\documentclass[reprint,aps,prb,twocolumn,superscriptaddress]{revtex4-1}
\usepackage{amssymb}
\usepackage{upgreek}
\usepackage{graphicx}
\usepackage{mathrsfs}
\usepackage{textcomp}
\usepackage[utf8]{inputenc}
\usepackage{color}
\usepackage{siunitx}
\usepackage{url}

\usepackage[utf8]{inputenc}
\usepackage{float}
\usepackage[paperwidth=210mm,paperheight=297mm,centering,hmargin=2cm,vmargin=2cm]{geometry}
\usepackage{lipsum}
\usepackage{circuitikz}
\usepackage{fancyhdr}
\usepackage{mathtools}
\usepackage{bm}
\usepackage[normalem]{ulem}
\usepackage{upgreek}
\DeclareGraphicsExtensions{.jpg,.pdf,.mps,.png}

\begin{document}

\title{Effects of Co substitution on the structural and magnetic properties of Sr(Ni$_{1-x}$Co$_x$)$_2$P$_2$}

\author{J. Schmidt}
\affiliation{Department of Physics and Astronomy, Iowa State University, Ames, IA 50011, USA}
\affiliation{Ames National Laboratory, Iowa State University, Ames, IA 50011, USA}
\author{G. Gorgen-Lesseux}
\affiliation{Department of Physics and Astronomy, Iowa State University, Ames, IA 50011, USA}
\affiliation{Ames National Laboratory, Iowa State University, Ames, IA 50011, USA}
\author{R. A. Ribeiro}
\affiliation{Department of Physics and Astronomy, Iowa State University, Ames, IA 50011, USA}
\affiliation{Ames National Laboratory, Iowa State University, Ames, IA 50011, USA}
\author{S. L. Bud'ko}
\affiliation{Department of Physics and Astronomy, Iowa State University, Ames, IA 50011, USA}
\affiliation{Ames National Laboratory, Iowa State University, Ames, IA 50011, USA}
\author{P. C. Canfield}
\affiliation{Department of Physics and Astronomy, Iowa State University, Ames, IA 50011, USA}
\affiliation{Ames National Laboratory, Iowa State University, Ames, IA 50011, USA}

\date{\today }

\pacs{1234}

\begin{abstract}

%Option A:

Although SrNi$_2$P$_2$ adopts the common ThCr$_2$Si$_2$ structure for $T\geq 325$ K, being in an uncollapsed tetragonal state rather than a collapsed tetragonal version, it is a special case for the ThCr$_2$Si$_2$ class: on cooling below 325 K it adopts a one-third collapsed orthorhombic phase where one out of every three P rows bond across the Sr layers. On the other hand, SrCo$_2$P$_2$ only exhibits the uncollapsed ThCr$_2$Si$_2$ structure from room temperature down to 1.8 K. Regardless of their low-temperature structures, neither SrNi$_2$P$_2$ nor SrCo$_2$P$_2$ manifests magnetic transitions down to 50 mK and 2 K, respectively. In this work we report the effects of Co substitution in Sr(Ni$_{1-x}$Co$_x$)$_2$P$_2$, which allows for tuning the transition between the one-third collapsed and the uncollapsed structure. We find a rapid decrease of the one-third collapsed structural transition temperature with increasing Co fraction, until reaching full suppression for $x \geq 0.1$. Substitution levels in the range $0.11\leq x\leq 0.58$ show no signs of any transition down to 1.8 K in the magnetization or resistance measurements in the range $1.8\ \leq T\leq 300\ \text{K}$. However, different magnetically ordered states emerge for $x\geq 0.65$, and disappear for $x\geq 0.99$, recovering the known paramagnetic properties of the parent compound SrCo$_2$P$_2$. These results are summarized in a phase diagram, built upon the characterization by energy dispersive x-ray spectroscopy, x-ray diffraction, temperature-dependent resistance, field- and temperature-dependent magnetization measurements done on single crystals with different Co fraction. Both the magnetic and structural properties are compared to other systems with  ThCr$_2$Si$_2$ structure that exhibit magnetic ordering and collapsed tetragonal transitions. The magnetic ordering and moment formation are well described by Takahashi's spin fluctuation theory of itinerant electron magnetism [Y. Takahashi, J. Phys. Soc. Jpn. \textbf{55}, 3553 (1986)].

\end{abstract}

\maketitle

\section{Introduction} 
\label{sec:Introduction}

Compounds with the ThCr$_2$Si$_2$ crystal structure have attracted abundant interest, offering a plethora of tunable magnetic, electronic, structural, and superconducting properties \cite{Szytula1994,Gati2020,Canfield2009,Ni2008a,Gati2012,Trovarelli2000,Li2019}. Among these compounds, some pnictides $A(TM)_2(Pn)_2$ (with $A$ generally an alkali metal, alkaline earth, or rare earth, $TM$ a transition metal and $Pn$ a pnictogen) display $Pn$-$Pn$ distances, across the $A$ layer, which are markedly smaller than others. This has been rationalized in terms of the bonding between $Pn$ atoms across the layers of $A$ atoms \cite{Hoffmann1985}, which results in a reduced $Pn$-$Pn$ distance as well as a smaller $c$-lattice parameter. The term collapsed tetragonal (cT) has been coined to identify those compounds, whereas those that do not present $Pn$-$Pn$ bonding are referred to as uncollapsed tetragonal (ucT). This terminology was extended to members of the CaKFe$_4$As$_4$ family of materials \cite{Iyo2016}, most commonly known as 1144, that can also display a half-collapsed tetragonal (hcT) phase in which As atoms bond across every other $A$ layer \cite{Yu2009,Kaluarachchi2017,Borisov2018}.

The $Pn$-$Pn$ distance in these systems has been proven to be highly tunable by different means such as applying pressure \cite{Kreyssig2008,Gati2012,Kaluarachchi2017,Borisov2018}, uniaxial strain \cite{Xiao2021}, and chemical substitution \cite{Jia2009}. This has allowed for experimental access to a collapsed tetragonal transition in which the $Pn$-$Pn$ bonds form or break, and for studies of how this impacts the mechanical, electronic, and magnetic properties of these materials. The cT and hcT transitions have been associated with changes from magnetic to nonmagnetic states \cite{Kreyssig2008,Canfield2009,Jia2011}, from superconducting to nonsuperconducting states \cite{Kaluarachchi2017,Gati2012} and, in addition, these transitions give rise to a remarkable pseudoelasticity in these materials \cite{Sypek,Bakst2018}, which allows them to achieve some of the highest maximum recoverable strains among metals \cite{Xiao2021}. 

It is worth highlighting the case of SrCo$_2$(Ge$_{1-x}$P$_x$)$_2$ for which one of the end members (SrCo$_2$Ge$_2$) has a cT structure, and the other (SrCo$_2$P$_2$) has a ucT structure. Despite both of these being paramagnetic at temperatures down to 2 K, intermediate compositions exhibit an emergence of ferromagnetic order \cite{Jia2011}. The magnetism in this series has been shown to occur simultaneously with the gradual breaking of the $X
$-$X$ bonds ($X =$ Ge, P), and therefore the two phenomena have been theoretically linked. In general, it has long been proposed \cite{Hoffmann1985} that there is a decrease of the $X$-$X$ $\sigma^*$ antibonding band population leading to the bond formation that takes place in a collapsed tetragonal transition, manifested as large anisotropic changes in the lattice parameters. Additionally, it has been suggested that itinerant ferromagnetism can be driven by a high density of states with strong antibonding character at the Fermi level \cite{Landrum2000}. In the case of  SrCo$_2$(Ge$_{1-x}$P$_x$)$_2$ ferromagnetism was therefore attributed to the gradual population and/or depopulation of the mentioned $\sigma^*$ band, potentially hybridized with the $3d$ bands of Co \cite{Jia2011}. Similar correspondence between collapsed tetragonal material and magnetic ordering have been observed in other systems as well \cite{Jia2009}.

The case of SrNi$_2$P$_2$ is special in terms of its bonding properties, compared to all the other $A(TM)_2(Pn)_2$ compounds. It presents a unique structure that is neither ucT nor cT, but one in which only one of every three P-P rows bonds across the Sr layers when cooled down below $325\ \text{K}$ \cite{Barth1997}. As such, this structure can be thought of as one-third collapsed across the Sr plane. Since the structure is now orthorhombic, SrNi$_2$P$_2$ exhibits a \textit{one-third collapsed orthorhombic} (tcO) structure. At room temperature and ambient pressure, those P atoms that are bonded are at a distance of $0.2451\ \text{nm}$ from each other, and those that are not are at a distance of $0.3282 \ \text{nm}$, giving an average P-P distance of $0.3005\ \text{nm}$ \cite{Keimes1997}. Additionally, SrNi$_2$P$_2$ can exhibit ucT and cT states under modest changes in temperature or pressure. At temperatures above 325 K, it presents a ucT structure, with a P-P distance of $0.3120\ \text{nm}$ reported for 373 K \cite{Keimes1997}; and upon applying only 4 kbars of hydrostatic pressure at room temperature, the structure collapses into a cT state \cite{Keimes1997}.

Motivated by the unique bonding properties of SrNi$_2$P$_2$ and the proposed relationship between bonding instability and the emergence of itinerant magnetic order in other related compounds, we explore the case of Sr(Ni$_{1-x}$Co$_x$)$_2$P$_2$. SrNi$_2$P$_2$ transitions from a ucT structure into a primarily tcO phase, but with 16.5\% of ucT remaining present at room temperature \cite{Xiao2021} which is relatively close to the transition temperature (325 K). In contrast, SrCo$_2$P$_2$ is known to remain uncollapsed down to at least 2 K \cite{Imai2015}. In addition, neither SrNi$_2$P$_2$ nor SrCo$_2$P$_2$ exhibits any magnetic transitions from room temperature to temperatures as low as 2 K \cite{Ronning2009,Sugiyama2015}. In this work we investigate the effects of Co substitution, $x$, which allows to tune the ucT $\leftrightarrow$ tcO transition temperature. Moreover, we report on magnetic ordering that occurs at intermediate values of $x$. Despite the superficial similarities to SrCo$_2$(Ge$_{1-x}$P$_x$)$_2$, and other related families such as Ca$_{1-x}$Sr$_x$Co$_2$P$_2$, we demonstrate that Sr(Ni$_{1-x}$Co$_x$)$_2$P$_2$ is inherently different from all the other previously explored examples and challenges the the proposed relationship between bonding and the emergence of itinerant magnetism in these systems.

\section{Experimental Details}
\label{sec:Experimental}

Single crystals of Sr$(TM)_2$P$_2$ were obtained by high-temperature solution growth method \cite{Canfield2020}
out of Sn flux. Pure metals were loaded into a $2\ \text{ml}$ alumina fritted Canfield crucible set \cite{CanfieldP.C.KongT.KaluarachchiU.S.2016,LSPCeramics}, and sealed under partial atmosphere of argon in a fused silica tube. Starting compositions of Sr$_{1.3}$(Ni$_{1-x}$Co$_x$)$_2$P$_{2.3}$Sn$_{16}$ were used for lower-doped ($x<0.7$) crystals and Sr$_{1.2}$(Ni$_{1-x}$Co$_x$)$_2$P$_{2}$Sn$_{20}$ for higher dopings. The ampoules were placed inside a box
furnace, held for 6 hours at $600\ ^{\circ}\text{C}$ before increasing to $1180\ ^{\circ}\text{C}$, dwelled for 24 hours to make sure the material was fully melted, and finally slowly cooled down over 100 hours to $950\ ^{\circ}\text{C} - 1000\ ^{\circ}\text{C}$, the temperature at which the excess Sn was decanted with the aid of a centrifuge. The decanting temperature was optimized in order to maximize homogeneity in composition, and lower temperatures down to $950\ ^{\circ}\text{C}$ were needed for compositions with $x<0.25$. Details on this are further described in Appendix A.

The Co concentration levels ($x$) were determined by energy dispersive x-ray spectroscopy (EDS) quantitative chemical analysis using an EDS detector (Thermo NORAN Microanalysis System, model C10001) attached to a JEOL scanning electron microscope (SEM). An acceleration voltage of
$22\ \text{kV}$, working distance of $10\ \text{mm}$, and takeoff angle of $35 ^{\circ}$ were used for measuring all standards and crystals with unknown composition. Single crystals of SrNi$_2$P$_2$ and of SrCo$_2$P$_2$ were used as standards for Sr, Ni, Co, and P quantification. The spectra were fitted using NIST-DTSA II Microscopium software \cite{Newbury2014}. The composition of each platelike crystal was measured at different positions on the crystal's face (perpendicular to $c$), as well as at different points across the edge of the crystal (along $c$). The average compositions and error bars were obtained from these data, accounting for both inhomogeneity and goodness of fit of each spectra. 

Powder x-ray diffraction (XRD) measurements were performed using a Rigaku MiniFlex II powder diffractometer with Cu K$\alpha$ radiation ($\lambda=1.5406\ \text{\AA}$). For each composition, a few crystals were finely ground to powder and dispersed evenly on a single-crystal Si zero-background holder, with the aid of a small quantity of vacuum grease. Intensities were collected for $2\theta$ ranging from $5^{\circ}$ to $100^{\circ}$, in step sizes of $0.01^{\circ}$, counting for 4 s at each angle. Rietveld refinement was performed on each spectrum using the GSAS II software package \cite{Dreele2014}. Refined parameters included but were not limited to phase fractions, lattice parameters, atomic positions and isotropic displacements.

DC magnetization measurements were carried out on a Quantum Design Magnetic Property Measurement System (MPMS classic) superconducting quantum interference device (SQUID) magnetometer (operated in the range $1.8\ \leq T \leq 350\ \text{K}$, $|H|\leq 50\ \text{kOe}$). Each sample was measured with the field applied parallel and perpendicular to the tetragonal $c$ axis. Most measurements were performed under zero-field cooling (ZFC) protocols, but field cooling (FC) protocols were also employed in some cases. The samples were glued on a Kel-F disk which was placed inside a plastic straw; the contribution of the disk to the measured magnetic moment was independently measured in order to subtract it from our results. The samples used for these measurements have a platelike shape, where the dimensions along the crystallographic $a$ and $b$ axes are at least ten times larger than the dimension along the $c$ axis. This results in demagnetization factors that are lower than 0.1 for the cases where the magnetic field is applied perpendicular to the $c$ axis. Hence, the product of the demagnetization factor and the magnetization values are much smaller than the applied magnetic fields for those cases. For the cases with the field applied parallel to the $c$ axis, despite having a larger demagnetization factor, the magnetization values correspond to less than 0.4\% of the applied magnetic field, so the demagnetization field can be neglected even for the maximum possible demagnetization factor of 1. 

Temperature-dependent resistance measurements were performed in closed-cycle cryostat Janis SHI-950, with phosphor–bronze wires (QT-36, LakeShore, Inc.) used on the probe. Temperature was measured by a calibrated Cernox-1030 sensor connected to a LakeShore 336 controler. The sample AC resistance was measured with LakeShore AC resistance bridges (Models 370 and 372), with a frequency of $17\ \text{Hz}$ and $3\ \text{mA}$ excitation current. In some cases, AC resistance was also measured using a Quantum Design Physical Property Measurement System (PPMS) using the ACT option, with the same frequency and excitation current. Only in-plane resistance was measured, using a standard four-contact geometry. Electrical contacts with less than $1.5\ \Omega$ resistance were achieved by spot-welding $25\ \mu\text{m}$ Pt wire to the samples, followed by adding Epotek H20E silver epoxy, and curing the latter for 1 hour at $120^{\circ}\text{C}$.

\section{Results}
\label{sec:Results}

\begin{figure}[tb]
 \includegraphics[width=\linewidth]{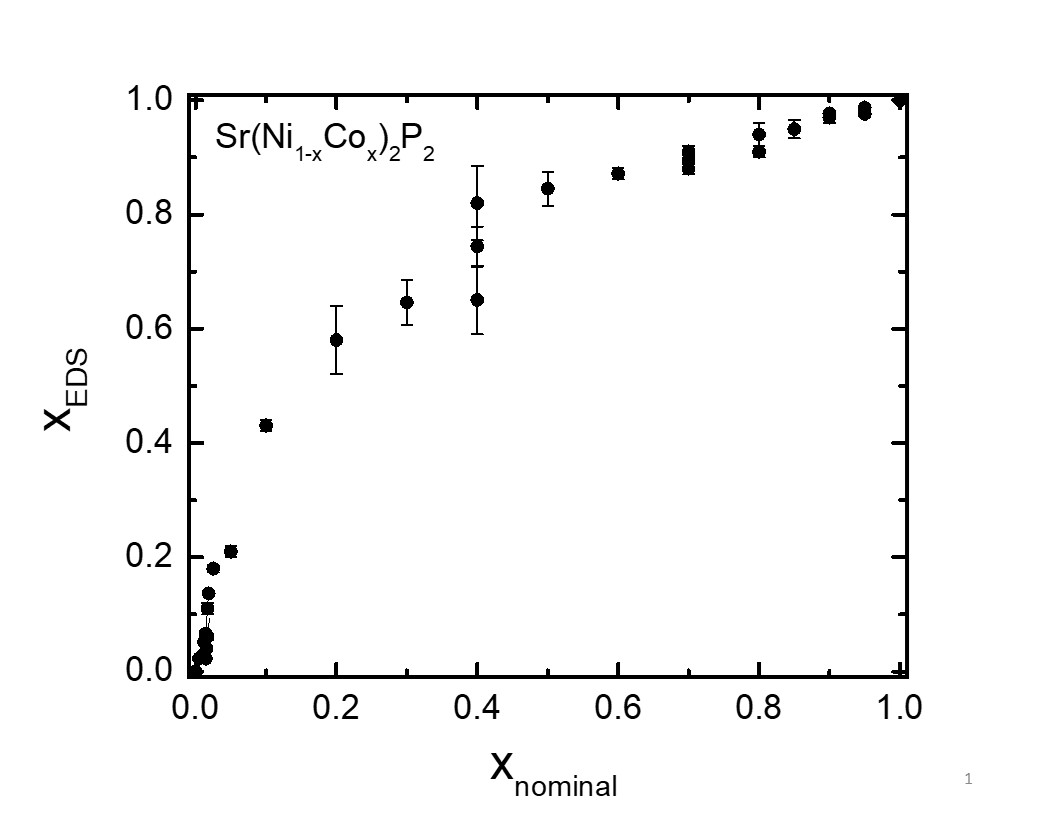}
 \caption{\footnotesize{Fraction of Co in Sr(Ni$_{1-x}$Co$_x$)$_2$P$_2$} determined by EDS as a function of the nominal Co fraction that was originally present in the high-temperature melt that crystals were grown out of.}
 \label{fig:EDS}
\end{figure}

The Co substitution determined by EDS, $x_{\text{EDS}}$, is shown in Fig. \ref{fig:EDS} as a function of the nominal Co fraction, $x_{\text{nominal}}$, that was originally used to create the high-temperature solution out of which the crystals were grown. For all the dopings explored in this work, $x_{\text{EDS}}$ was larger than $x_{\text{nominal}}$. If any, inhomogeneities were not detected when measuring at different positions on the platelike crystal's surface facets (perpendicular to the $c$ axis) but rather for different points across the edge of the crystal (along the $c$ axis). Although these were minimized by adjusting the decanting temperature in the crystal growth procedure (as detailed in Appendix A), some degree of inhomogeneity can still be detected as seen in the error bars in the samples with $0.2 \leq x_{\text{nominal}} \leq 0.5$, or $0.58 \leq x_{\text{EDS}} \leq 0.85$ (Fig. \ref{fig:EDS}). More details on this can be found in Appendix A. It should be mentioned that, given the sensitivity of the magnetic properties to $x$ when approaching pure SrCo$_2$P$_2$, samples with small differences in $x$ were studied. In particular, the $x_{EDS}$ measured on two of the samples was the same (0.98) within the resolution of the measurement (0.01), even though they were obtained from different batches and show differences in the analyzed properties in this work, and are thus both reported separately in this work. For simplicity, in the rest of this paper the bare symbol $x$ will be used to refer to the Co fraction determined by EDS.

\begin{figure}[tbp]
\centering
\includegraphics[width=\linewidth]{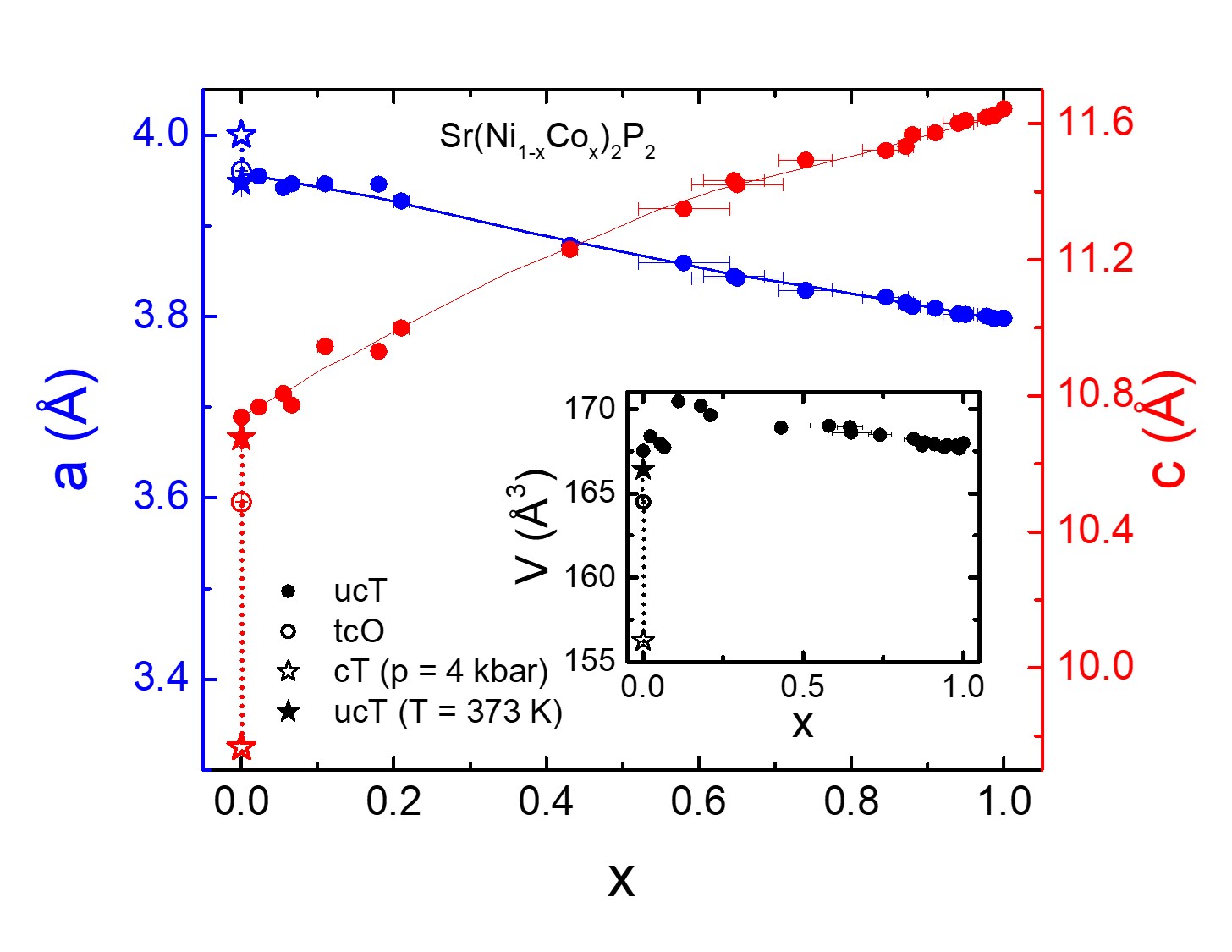}
\caption{\footnotesize{ $a$- and $c$-lattice parameters (main panel) and unit cell volume (inset) at room temperature as a function of the Co fraction, $x_{EDS}$}. The lattice parameters and volume of the cT phase under a pressure \cite{Keimes1997} of $4\ \text{kbar}$ are shown with open star symbols, and those of the ucT phase at a temperature of 373 K \cite{Keimes1997} are shown with solid star symbols. The ucT and cT symbols (stars) are connected with a dotted line so as to help illustrate the respective changes in $a$- and $c$-lattic parameters.}
\label{fig:latticeparam}
\end{figure}

Figure \ref{fig:latticeparam} shows the $a$- and $c$-lattice parameters obtained by Rietveld refinement of the powder XRD patterns measured at room temperature as a function of $x$. The tcO and ucT phases are distinguishable from each other due to the fact that they belong to different space groups. Particularly, the tcO (space group $Immm$) phase exhibits peaks that are forbidden for the ucT (space group $I4/mmm$) due to the former's reduced symmetry. When performing the refinement, coexistence of both phases was allowed, and lattice parameters of both phases were refined. The open symbols in the figure correspond to the lattice parameters of the tcO phase, whereas the solid symbols correspond to those of the ucT phase. As shown, coexistence of both phases at room temperature was found for pure SrNi$_2$P$_2$, consistent with previous reports which indicate that the two structures are coherently connected in a two-phase single crystal \cite{Xiao2021}. On the other hand, all the Co-doped crystals ($x\geq 0.023$) resulted in single-phase ucT material at room temperature with no detectable peaks indicative of tcO phase within experimental resolution. This is consistent with the expected role of Co in  suppressing the tcO transition, stabilizing the ucT phase at room temperature (see Figs. \ref{fig:huge_plot} and \ref{fig:phase_diagram} below). 

Figure \ref{fig:latticeparam} also shows that, at $x=0$, the ucT phase has a significantly larger $c$-lattice parameter than the tcO phase, by $\Delta c(0) \sim 0.25\  \text{\AA}$. This jump is comparatively smaller than the one observed for the pressure-induced transition at $4\ \text{kbars}$ which goes from $c=10.40\ \text{\AA}$ (in the tcO phase) to $c=9.76\ \text{\AA}$ (in the cT phase) \cite{Keimes1997}. An additional significant increase of $c$ is observed for the ucT phase due to further Co substitution, with a change of $0.9\ \text{\AA}$ across the full composition range. Even though a relatively small amount of Co is needed to fully stabilize the uncollapsed state at ambient temperature, there is a clearly enhanced sensitivity of $c$ with $x$ beyond the structural transition. This change in $c$-lattice parameter is comparable to that seen between BaNi$_2$P$_2$ \cite{Keimes1997} and BaCo$_2$P$_2$ \cite{Imai_2017} ($\sim 0.6\ \text{\AA}$) \cite{Imai_2017}, but is significantly larger than that seen in other 122 compounds \cite{Mewis1980,HOFMANN1984,REEHUIS1998}. 

Much smaller changes were observed in the $a$-lattice parameter between tcO and ucT phases at $x=0$ [it should be noted that the value used for the tcO phase was the average of the $a=3.9601(5)$ and $b/3=3.9591(6)$ lattice parameters]. However, a large decrease of $a$ can be observed (blue solid symbols) with increasing $x$. This effect compensates the decrease of $c$, resulting in a nearly constant unit cell volume $V$ of the ucT phase across the full composition range, as shown in the inset in Fig. \ref{fig:latticeparam}. In fact the variations in $V$ are of 1.8\%, which is very close to the percentage difference in volume of $1.6\%$ between Ni and Co pure \textit{fcc} metals \cite{Panday2011}, and contrasts with the 8.0\% change in $c$ and  3.9\% change in $a$. This compensation of lattice parameter changes is not uncommon, and often occurs in other systems.

\begin{figure*}
\centering
\includegraphics[width=\linewidth]{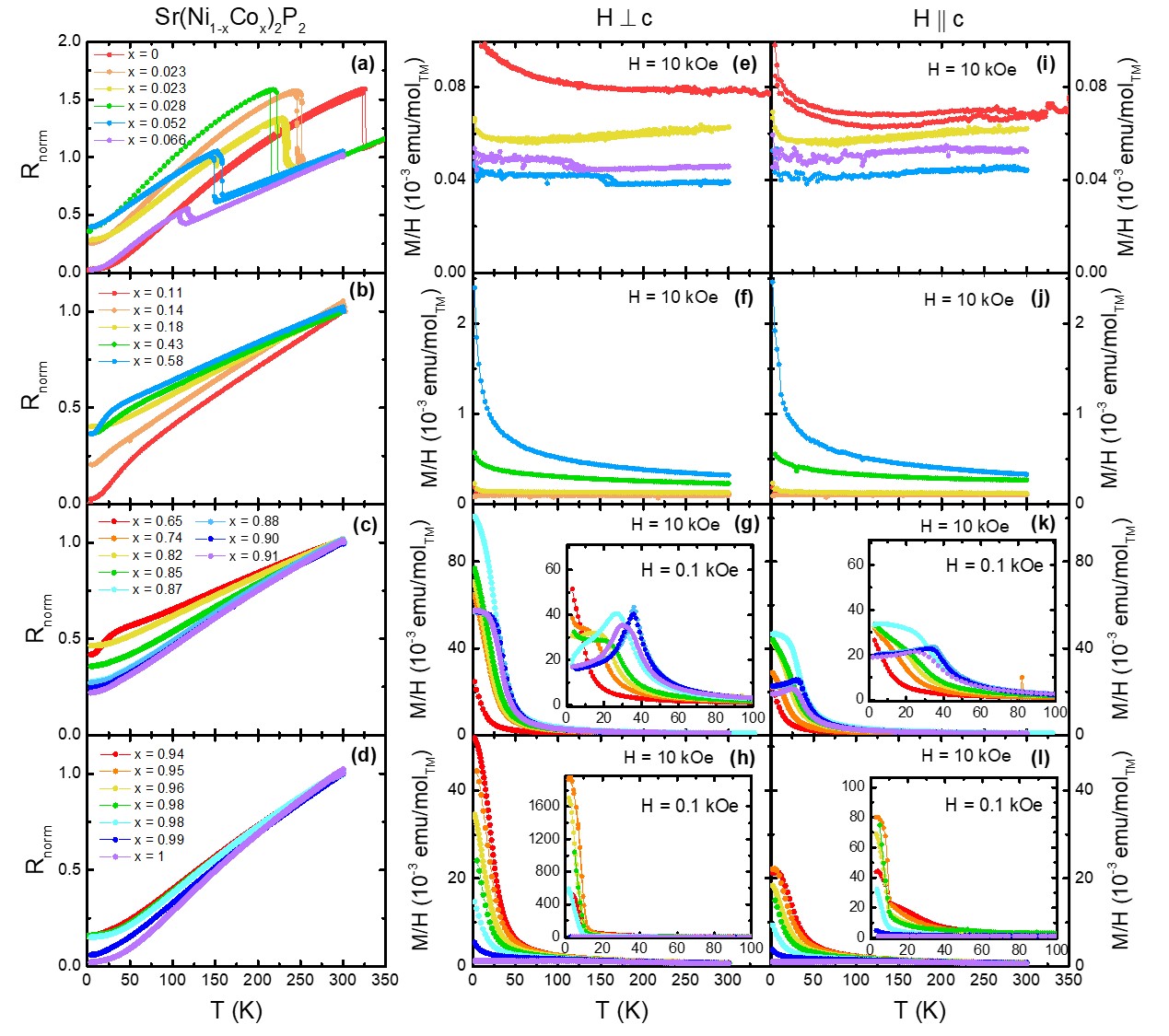}
\caption{\footnotesize{(a)-(d) Temperature-dependent resistance normalized to its value at $300\ \text{K}$, $R(T)/R(300\ \text{K})$, measured upon cooling and warming for the samples that exhibit signatures of collapsed tetragonal transition, no transition, AFM transition and FM transition, respectively. For (a) $R(T)$ for $x=0$ is normalized to the value of $R($350 K$)$ of $x=0.028$. (e)-(l) ZFC temperature dependent magnetization divided by the applied field, $M(T)/H$, for samples in same substitution regions as for (a)-(d). Panels (e)-(h) correspond to fields applied perpendicular to the $c$-axis, whereas (i)-(l) correspond to fields applied along $c$. Insets in panels (g), (h), (k) and (l) correspond to lower fields ($0.1\ \text{kOe}$) than those used for the data shown in the main panels ($10\ \text{kOe}$).}}
\label{fig:huge_plot}
\end{figure*} 

\begin{figure}
\centering
\includegraphics[width=\linewidth]{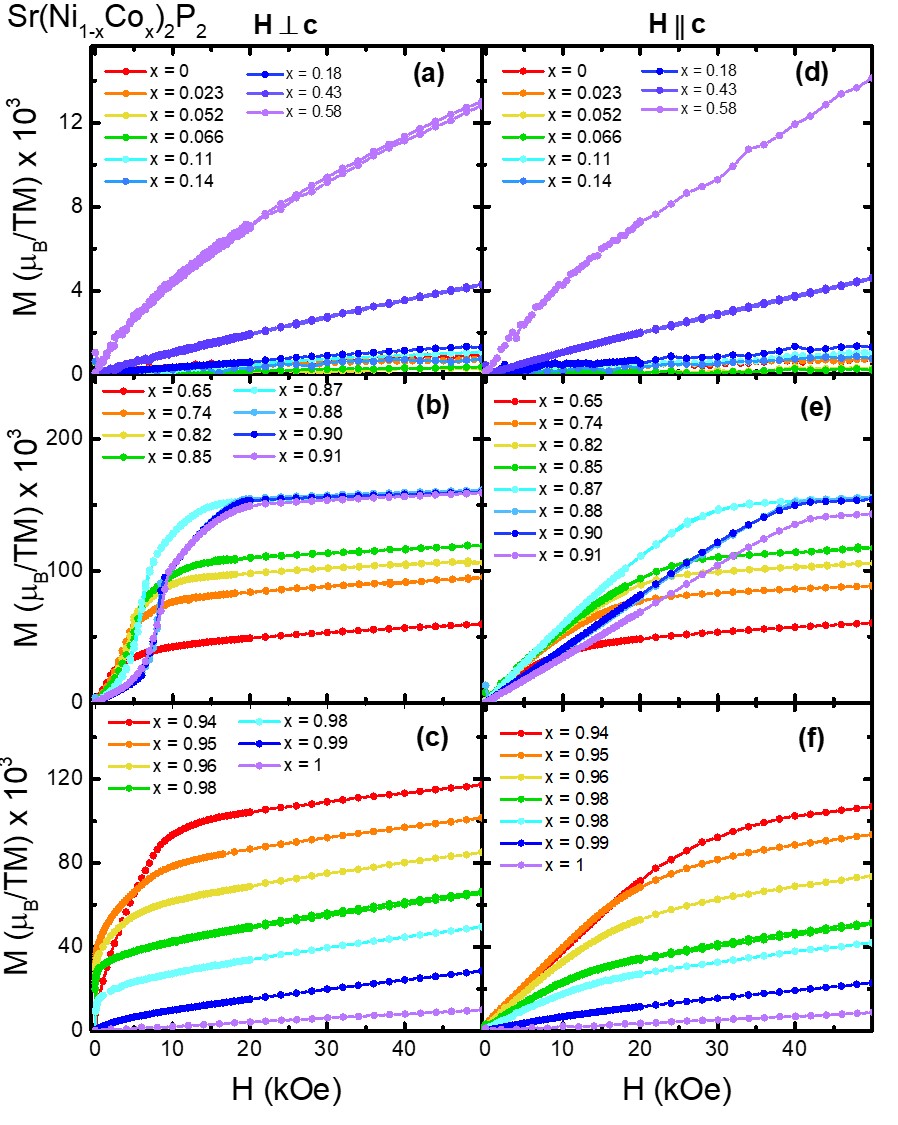}
\caption{\footnotesize{Field-dependent magnetization data, in units of $\mu_B$ per transition metal atom, for the applied field perpendicular to $c$ (panels (a)-(c)) as well as parallel to $c$ (panels (d)-(f)). Panels (a) and (d) correspond to samples with no magnetic order down to $1.8\ \text{K}$; (b) and (e) to samples with AFM order; (c) and (f) to samples with FM order.}}
\label{fig:MH_complete}
\end{figure}

Temperature-dependent resistance and anisotropic magnetization data were collected for all samples.  Figure \ref{fig:huge_plot} presents these data for various ranges of $x$. $M/H$ was obtained for the magnetic field applied perpendicular to the $c$ axis (parallel to the basal $a$-$b$ plane) shown in Fig. \ref{fig:huge_plot}(e)-(h), as well as parallel to the crystallographic $c$ axis shown in Fig. \ref{fig:huge_plot}(i)-(l), following a ZFC protocol. The $M(T)/H$ is expressed in units of emu per mole of transition metal, by means of
\begin{equation}
    M/H(\text{emu}/\text{mol}_{TM})=\frac{M(\text{emu})}{2n_{f.u.}(\text{mol}) H(\text{Oe})},
\end{equation}
where $M$ is the measured magnetic moment, $H$ is the applied field and $n_{f.u.}$ is the number of moles of formula units present in the measured crystal, which is multiplied by 2 to account for the fact that there are two transition metals per formula unit of Sr$TM_2$P$_2$. In addition, anisotropic field-dependent magnetization data were collected for $T = 2\ \text{K}$; these data are shown in Fig. \ref{fig:MH_complete}. From all of these sets of data we are able to assemble a $T$-$x$ phase diagram (shown in Fig. \ref{fig:phase_diagram}) for the Sr(Ni$_{1-x}$Co$_x$)$_2$P$_2$ system.  In the sections below we will examine different regions in detail.  For small $x$-values ($x < 0.10$) we see a monotonic suppression of the tcO transition temperature.  For intermediate $x$ values ($0.10 < x < 0.50$) no signatures of any phase transition are found for $2\ \text{K}< T < 300\ \text{K}$.  For higher $x$ values ($0.58 < x < 0.99$) we find signatures of magnetic, first antiferromagnetic (AFM) and ultimately ferromagnetic (FM), transitions forming a peaked-dome-like structure. The latter can be regarded as weak itinerant ferromagnetism, since the saturation magnetization values correspond to fractions of a $\mu_B$ per transition metal atom, and the ordering temperatures are low (see further discussion below). At the highest $x$-value region (x = 0.99, 1.00) there is again no detectable phase transition for $T > 1.8\ \text{K}$.  In the following sections we will discuss these regions is greater detail.

For the compositions shown in Fig. \ref{fig:huge_plot}(a), a sharp steplike feature in the temperature-dependent resistance can be identified as a fingerprint of the tcO $\leftrightarrow$ ucT structural transition temperature $T_S$. For some of these compositions, temperature-dependent magnetization also revealed a more subtle step when the field was applied perpendicular to the $c$ axis (Fig. \ref{fig:huge_plot}(e)), although it was not clearly observed for all of them, since the magnetic signal of the structural transition for these $x$ values was small and becoming comparable to the noise. $T_S$ decreases with increasing $x$, dropping below 2 K between $x=0.066$ and $x=0.11$.

For samples with $x>0.1$ no such step-like feature is observed for resistivity [Figs. \ref{fig:huge_plot}(b)-(d)] or magnetization measurements [Figs. \ref{fig:huge_plot}(f)-(h) and (j)-(d)], indicating that the ucT phase is stable down to $1.8\ \text{K}$ for $0.11 \leq x \leq 0.58$. All this is, once more, consistent with the fact that Co stabilizes the uncollapsed state. More specifically, Figs. \ref{fig:huge_plot}(b), (f) and (j) show no transition whatsoever for $0.11\leq x \leq 0.58$. A broad shoulderlike, crossover feature in the resistance can be observed for some of the compositions in Fig. \ref{fig:huge_plot}(b). The origin or possible significance of this feature is not yet understood, and no correlation to the magnetization measurements of Figs. \ref{fig:huge_plot}(f) and \ref{fig:huge_plot}(j) seems to be found. Figures \ref{fig:MH_complete}(a) and \ref{fig:MH_complete}(d) show the $M(H)$ isotherms for 1.8 K for samples with $x\leq0.58$ for fields parallel and perpendicular to $c$ axis. Both directions of applied field are consistent with nonmagnetically ordered states with the samples becoming increasingly more polarizable with increasing $x$.

\begin{figure}[H]
 \centering
 \includegraphics[width=\linewidth]{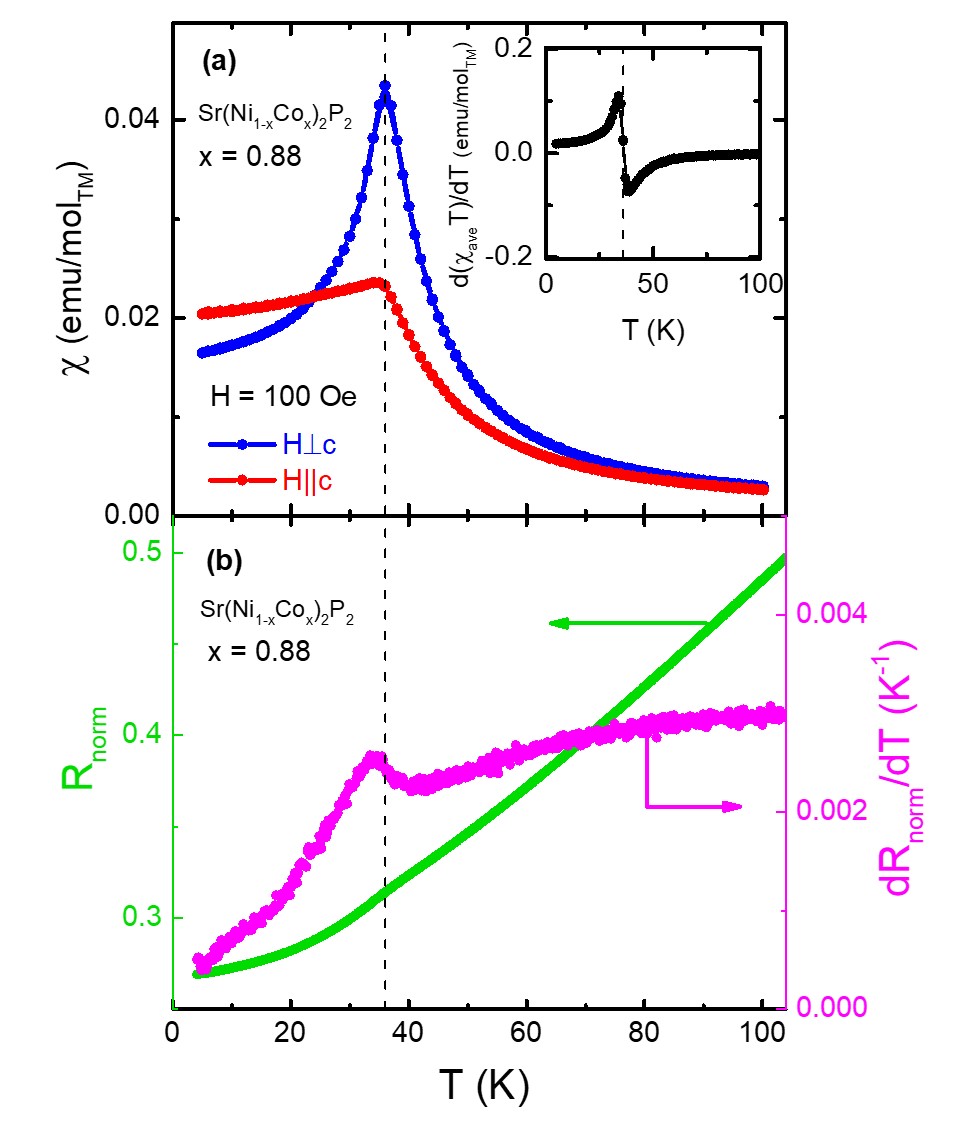}
 \caption{\footnotesize{(a) Temperature-dependent DC magnetic susceptibility $\chi$ for Sr(Ni$_{0.12}$Co$_{0.88}$)$_2$P$_2$ measured with an applied field of $100\ Oe$ perpendicular (blue) and parallel (red) to the $c$ axis. Inset: Temperature derivative of $\chi T$ as a function of temperature. (b) Temperature-dependent resistance $R_{\text{norm}}$ normalized to its room-temperature value (green, left axis), and its derivative $dR_{\text{norm}}/dT$} (magenta, right axis). AFM transition is marked with dashed lines.}
 \label{fig:MT_RT_70n_example}
\end{figure}

The magnetization data displayed in Figs. \ref{fig:huge_plot}(g) and \ref{fig:huge_plot}(k), as well as Figs. \ref{fig:MH_complete}(b) and \ref{fig:MH_complete}(e), suggest that AFM behavior is observed at low temperatures for the compositions ranging between $x=0.65$ and $x=0.91$. This is more clearly noted in the temperature-dependent magnetization measured at $H=0.1\ \text{kOe}$ (Fig. \ref{fig:huge_plot} insets) than at $H=10\ \text{kOe}$ (Fig. \ref{fig:huge_plot}, main panels). Figure \ref{fig:MT_100Oe_alltogether} of Appendix B shows in more detail the measurements done at $H=0.1\ \text{kOe}$ for both directions following a ZFC protocol, as well as following a FC protocol for $H\perp c$ for the samples with $0.58 \leq x \leq 0.91$. The magnetization results at $0.1\ \text{kOe}$ allow for the determination of the magnetic transition temperature $T_N$, plotted on the $T-x$ phase diagram shown in Fig. \ref{fig:phase_diagram}. Fig. \ref{fig:MT_RT_70n_example}(a) shows the case of $x=0.88$ as an example. Since the applied field of 0.1 kOe is within the range in which $M$ is proportional to $H$, the DC magnetic susceptibility can be obtained as
\begin{equation}
    \chi_j=\frac{M_j(H=0.1\ \text{kOe})}{0.1\ \text{kOe}},
\end{equation}
where $j$ index stands for the fact that the field can be applied perpendicular ($j=\perp$) or parallel ($j=||$) to the $c$ axis. The polycrystalline average of the susceptibility can be defined as
\begin{equation}
    \chi_{\text{ave}}=\frac{2\chi_{\perp}+\chi_{||}}{3},
\end{equation}
given that these systems present tetragonal symmetry.
In the inset, $d(\chi_{\text{ave}} T)/dT$ is plotted as a function of temperature, displaying a clear feature at $T=36\ \text{K}$. According to Ref. [\citenum{Fisher1962}], this magnitude should scale with the specific heat $C_p(T)$ for $T$ close to $T_N$, indicating the presence of an antiferromagnetic phase transition at $T_N=36\ \text{K}$, which was determined as the temperature at which the slope of  the function $f(T)=d(\chi_{\text{ave}} T)/dT$ is maximum. $d(\chi_{\text{ave}} T)/dT$ data for all the compositions are shown in detail in Fig. \ref{fig:MT_100Oe_alltogether} of Appendix B.  As can be appreciated in the insets of Figs. \ref{fig:huge_plot}(g) and \ref{fig:huge_plot}(k), as well as in Fig. \ref{fig:MT_100Oe_alltogether}, the transitions observed for $x=0.64-0.85$ are significantly broader than for $x=0.87-0.91$. The former compositions are those with larger degree of composition inhomogeneity, as shown in Fig. \ref{fig:EDS}, which could explain the increase in the transition width. Moreover, due to the presence of a local maximum in the phase diagram of Fig. \ref{fig:phase_diagram}, the sensitivity of the apparent width of $T_N$ to $x$ for the range $x=0.87-0.91$ is lower than for $x=0.64-0.85$, where $T_N(x)$ has a larger slope. This implies that a given spread of compositions in an individual sample has less impact on the spread of the transition temperatures for $x=0.87-0.91$.

The resistance measurements for the compositions in the $0.87\leq x\leq0.91$ range exhibit a loss of spin disorder scattering. Since this feature is too subtle to be observed in Fig. \ref{fig:huge_plot}(c), the case of $x=0.88$ is presented as an example in Fig. \ref{fig:MT_RT_70n_example}(b), where the green curve corresponds to the resistance normalized to its room-temperature value, and magenta curve corresponds to its derivative $dR_{\text{norm}}/dT$. The latter can be used to identify $T_N$, since its magnetic contribution should vary like the magnetic specific heat \cite{Fisher1968}.

The possibility of having a spin glass instead of an AFM state for some of the compositions in the $0.65\leq x\leq0.91$ range cannot be completely ignored. However, this is a less likely scenario due to the following reasons: (1) the fact that there is no measurable difference (irreversibility) between the ZFC and FC magnetization measurements as shown in Fig. \ref{fig:MT_100Oe_alltogether} of Appendix B, (2) the appearance of metamagnetic transitions in the $M(H)$ behavior, and (3) the observation of loss-of-spin-disorder scattering in some of the resistance curves as an indication of some long-range magnetic ordering being established. In order to entirely rule out the possibility of a spin glass instead of a long-range AFM state, neutron diffraction experiments should be carried out on these samples, which are outside of the scope of the current work. Other factors regarding this topic are addressed in the Discussion section.

Figures \ref{fig:huge_plot}(h) and \ref{fig:huge_plot}(l), as well as Fig. \ref{fig:MH_complete}(c), reveal a behavior consistent with a magnetically ordered state with a FM component at low temperatures. The criterion of Ref. [\citenum{Fisher1962}] does not hold to determine a FM transition temperature $T_C$.  Unfortunately, the loss-of-spin-disorder scattering in the electrical transport measurements was not resolvable for samples that revealed FM-like behavior. Even though loss-of-spin-disorder scattering may still occur in these samples upon cooling below $T_C$, the feature is too weak to be detected within our experimental resolution.

\begin{figure}[H]
 \centering
 \includegraphics[width=\linewidth]{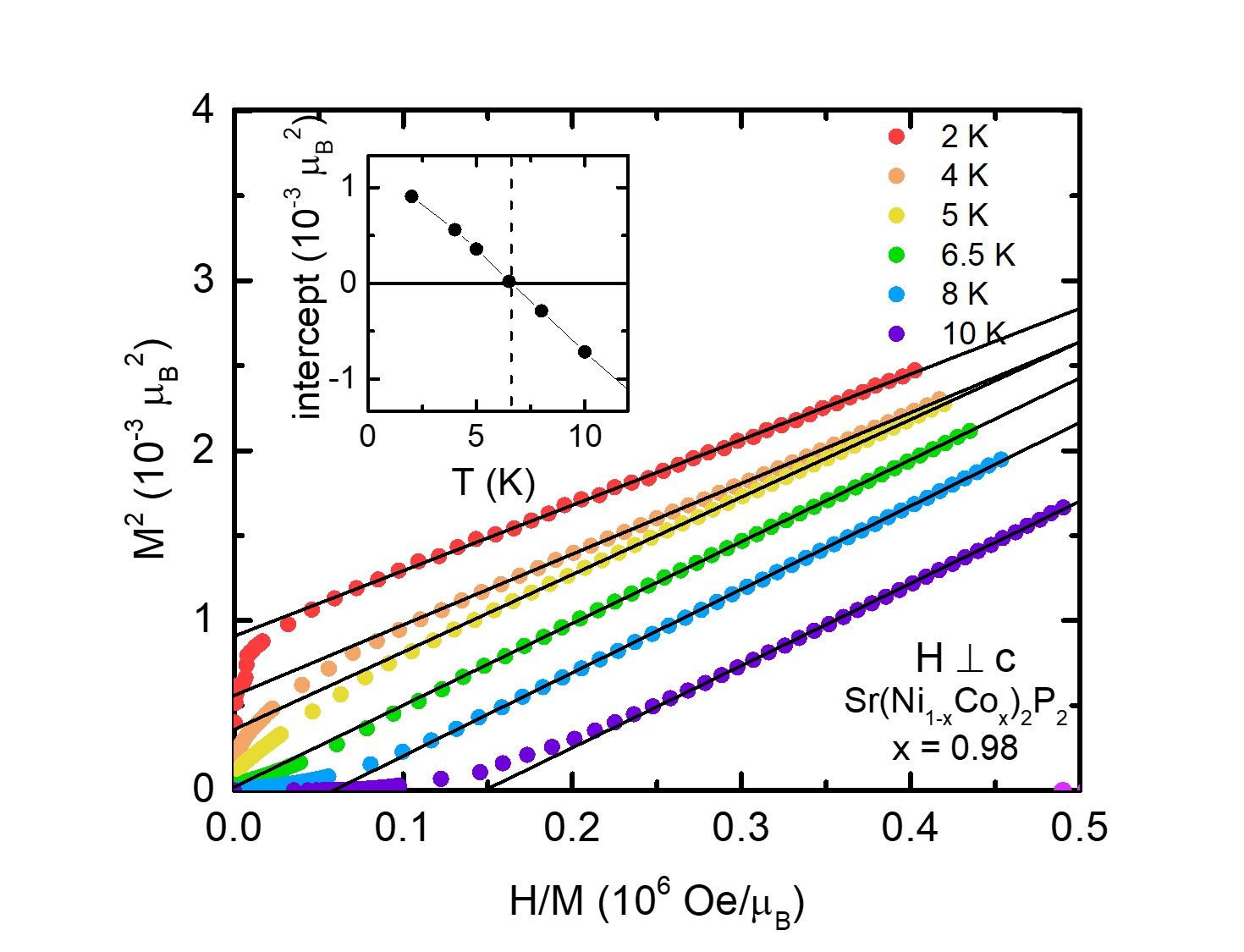}
 \caption{\footnotesize{ Main panel: $M^2$ as a function of $H/M$ (Arrott plot) at different constant temperatures for a sample with $x=0.98$, and their corresponding linear fits (solid lines). Inset: Value of the intercept obtained from the linear fits for the different temperatures.}}
 \label{fig:Arrot}
\end{figure}

In order to determine $T_C$ values, Arrott plots \cite{Arrott1957} were constructed for the samples that exhibit a FM transition. For this, field-dependent magnetization was measured at different temperatures, keeping $H$ perpendicular to the $c$ axis. Figure \ref{fig:Arrot} shows $M^2$ plotted as a function of $H/M$ for a sample with $x=0.98$, displaying a linear dependence as expected for a system that behaves according to the mean-field theory. A linear regression can be used to fit the higher field behavior observed for all the temperatures measured. This linear behavior can be extrapolated to zero field, in order to determine whether $T_C$ is higher or lower than the temperature at which the measurement was performed. The inset of the figure shows that the intercepts of the linear fits follow a clear trend with temperature, being positive for $T<6.5\ \text{K}$ and negative for $T>6.5\ \text{K}$. A linear fit was performed for those points closest to the $x$-axis intercept. The $x$-axis intercept was taken as the best estimate for $T_C$, and the corresponding error was derived by propagation of the errors of the linear fit parameters, obtaining a result of $T_C=6.6\pm0.4 \ \text{K}$. Similar results are shown in the Appendix C for other compositions. For some of the selected samples, as the ones shown in Figs. \ref{fig:Arrot} and \ref{fig:arrot_TM174}, the field-dependent magnetization measured at $2\ \text{K}$ exhibits a small steplike feature at $H\sim 200\ \text{Oe}$ due to the superconducting signal of Sn impurity, which is emphasized when plotting $M^2$ in an Arrott plot. Nevertheless, a change from negative concavity of the curves for temperatures below $T_C$ (red, orange, and yellow curves) to positive concavity above $T_C$ (green, blue, and purple curves) can be appreciated at low fields, as expected in Arrott plots \cite{Arrott1957}.

\begin{figure}[H]
 \centering
 \includegraphics[width=\linewidth]{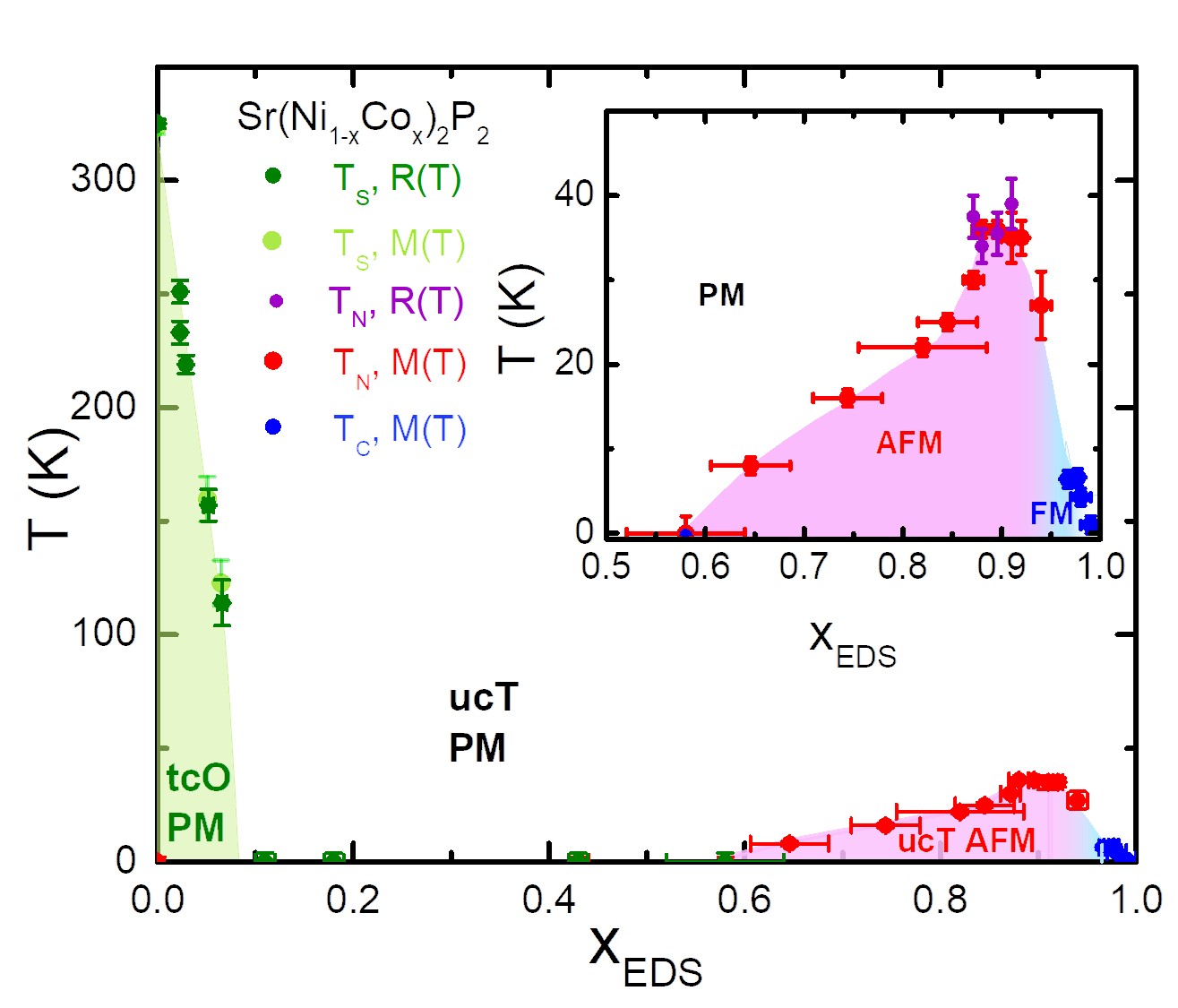}
 \caption{\footnotesize{Main panel: Phase diagram including all the critical temperatures determined: $T_S$ (green symbols from resistance measurements, and light green symbols from magnetization measurements), $T_N$ (red symbols from magnetization measurements and purple symbols from resistance measurements) and $T_C$ (blue symbols from magnetization measurements) as a function of Co fraction, $x$. Inset: enlarged view of the magnetic region of the phase diagram. Different colors are used to indicate the different phases: green for tcO paramagnetic, white for ucT paramagnetic, pink for ucT AFM, and light-blue for ucT FM.}}
 \label{fig:phase_diagram}
\end{figure}

All the critical temperatures determined, $T_S$, $T_N$ and $T_C$, are shown in the temperature-composition phase diagram depicted in Fig. \ref{fig:phase_diagram} (above). This summarizes the effects of varying $x$ from $0$ to $1$ in Sr(Ni$_{1-x}$Co$_x$)$_2$P$_2$. For low concentrations of Co it can be seen how the tcO $\longleftrightarrow$ ucT transition temperature decreases, and that the tcO state is fully suppressed for $x>0.1$. Moreover, upon further substitution ($x>0.58$) an AFM-like state emerges, and its transition temperature reaches a maximum $T_N\sim 36\ \text{K}$ for $x\sim0.9$. Beyond this doping, the transition temperature rapidly decreases, and the transition becomes ferromagnetic in nature. Finally, as Co concentration approaches $x=1$, the weak ferromagnetism is suppressed until reaching the nonmagnetically ordered (i.e. paramagnetic) SrCo$_2$P$_2$.

Systematic changes of the FM transition temperature with $x$, as shown in blue in Fig. \ref{fig:phase_diagram}, suggest that ferromagnetism is intrinsically coming from Sr(Ni$_{1-x}$Co$_x$)$_2$P$_2$, rather than from a possible secondary phase with a fixed composition. However, the results presented are not sufficient to conclude that the samples as a whole are FM. Since the change from AFM to FM-like behavior occurs suddenly in a small range of $x$, small inhomogeneities of $x$ within the EDS measurement resolution could lead to part of the sample being FM while another part being AFM. But even if that were not the case and the whole sample had homogeneous magnetic properties, a magnetic order with coexisting FM and AFM components cannot be ruled out for the compositions with $0.95<x< 0.99$, based on the measurements presented here. For these reasons, despite identifying these samples as FM for simplicity, all that can be stated is that they present a nonzero FM component. Further measurements, such as neutron diffraction, will be needed to further refine our understanding of this state. 

\section{Discussion}
\label{sec:Discussion}

\begin{figure}[H]
 \centering
 \includegraphics[width=\linewidth]{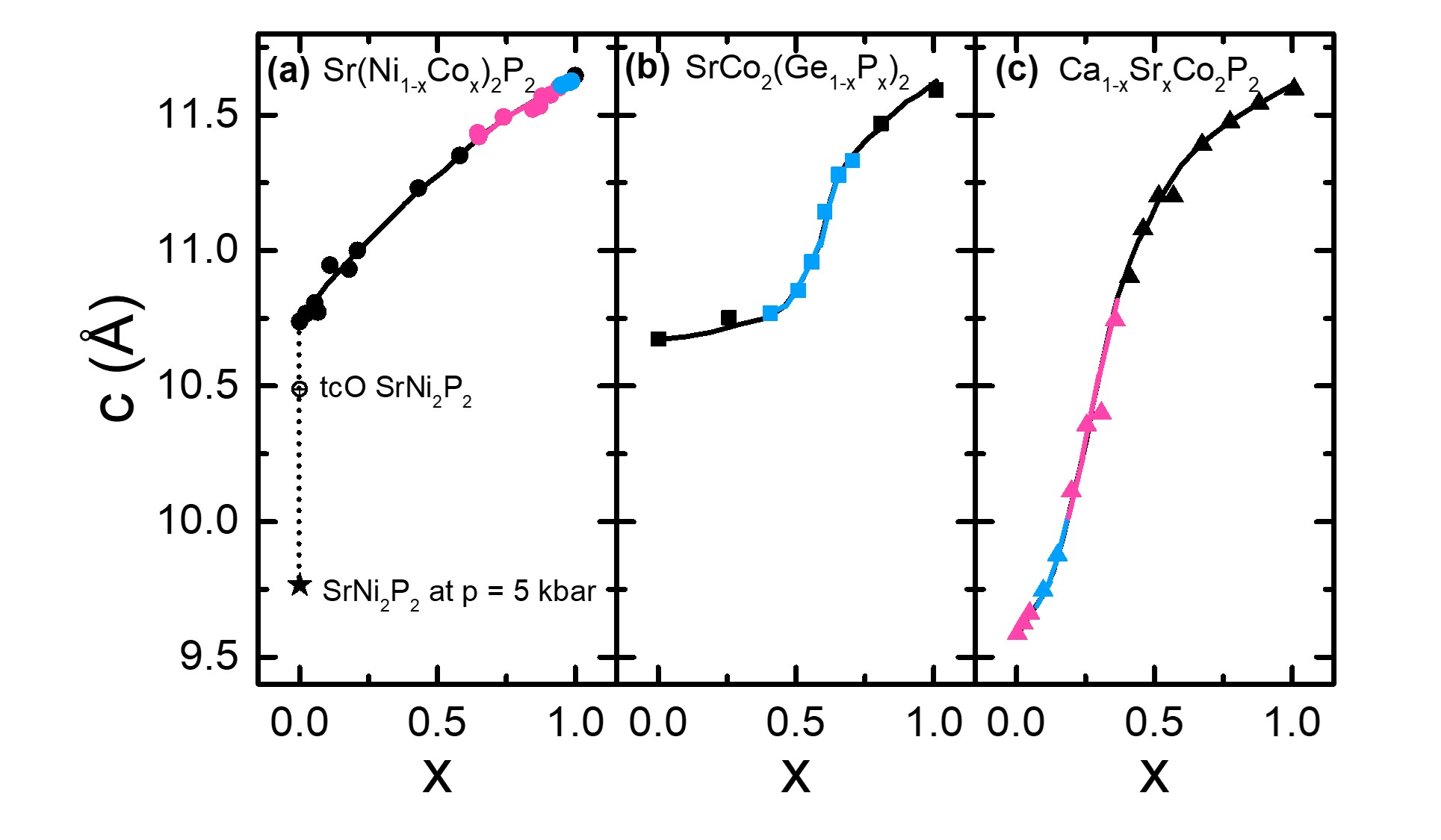}
 \caption{\footnotesize{$c$-lattice parameter as a function of $x$ for: (a) Sr(Ni$_{1-x}$Co$_x$)$_2$P$_2$, (b)  SrCo$_2$(Ge$_{1-x}$P$_x$)$_2$ \cite{Jia2011}, and (c) Ca$_{1-x}$Sr$_x$Co$_2$P$_2$ \cite{Jia2009}. The compositions that exhibit FM behavior are marked in blue, and those that exhibit AFM behavior are marked in magenta. Additionally, the $c$-lattice parameter for SrNi$_2$P$_2$ in the tcO phase (open red circles) and in the cT phase under pressure \cite{Keimes1997} (solid red star) are included.}}
 \label{fig:comp_lattice_param}
\end{figure}

As already mentioned, Fig. \ref{fig:latticeparam} shows that the Ni-Co substitution in Sr(Ni$_{1-x}$Co$_x$)$_2$P$_2$ leads to a significant change in the $c$-lattice parameter not only at the tcO transition but also across the whole ucT region. Figure \ref{fig:comp_lattice_param} shows the $c$-axis evolution of Sr(Ni$_{1-x}$Co$_x$)$_2$P$_2$ compared to that of SrCo$_2$(Ge$_{1-x}$P$_x$)$_2$ \cite{Jia2011} and Ca$_{1-x}$Sr$_x$Co$_2$P$_2$ \cite{Jia2009}. Each of these systems has a cT or tcO transition, and each of these systems has a magnetic order that evolves with substitution. 

Qualitatively the three systems have very similar behavior as $x$ is reduced from 1.0 to 0.7. Whereas one may anticipate a shrinking of the unit cell volume (and $c$-lattice parameter) for the substitution of Ca for Sr, a similar shrinkage for the substitution of Ge for P is counterintuitive \cite{Mills2000,Goetsch2012}. Given the similarity of these three systems for $0.7<x<1.0$, it is reasonable to argue that the effects the P-P bond instability persist to this range and may the common root of the observed changes in $c$-lattice parameter. Over a wider substitution range, the similarities disappear since the nature of the collapse and the compositions at which it occurs differ for each system. It is relevant to point out that the Sr(Ni$_{1-x}$Co$_x$)$_2$P$_2$ system is unique in that there is no transition to a cT state induced by doping, in contrast to SrCo$_2$(Ge$_{1-x}$P$_x$)$_2$ and Ca$_{1-x}$Sr$_x$Co$_2$P$_2$. Instead, the former only undergoes a transition into a tcO state very close to the SrNi$_2$P$_2$ end, and needs an additional applied pressure of $4\ \text{kbars}$ in order to decrease the value of $c$ to $9.76\ \text{\AA}$ and fully collapse into a cT state \cite{Keimes1997}. 

All three systems included in Fig. \ref{fig:comp_lattice_param} exhibit magnetic ordering upon perturbing the Stoner enhanced SrCo$_2$P$_2$. The blue points correspond to FM order, and the magenta points to AFM order. In the case of SrCo$_2$(Ge$_{1-x}$P$_x$)$_2$, the ferromagnetism has been attributed to the gradual population and/or depopulation of the $\sigma^*$ band, potentially hybridized with the $3d$ bands \cite{Jia2011}. It has been suggested that itinerant ferromagnetism can be driven by strong antibonding character of the bands at the chemical potential \cite{Landrum2000}.
FM order in the SrCo$_2$(Ge$_{1-x}$P$_x$)$_2$ is limited to the compositions $0.3<x<0.7$ [see Fig. \ref{fig:comp_lattice_param}(b), blue points], which corresponds to the region with the fastest change of the $c$-lattice parameter as consequence of a gradual breaking and/or formation of the bonds. In the case of Ca$_{1-x}$Sr$_x$Co$_2$P$_2$ magnetic ordering also appears in the high-slope region associated to the cT transition, and extends to the pure CaCo$_2$P$_2$ compound. In addition, the nature of the magnetic order in Ca$_{1-x}$Sr$_x$Co$_2$P$_2$ is FM for some compositions [see panel (c), blue points], and AFM for others (magenta points). Finally, Sr(Ni$_{1-x}$Co$_x$)$_2$P$_2$ presents both FM and AFM order, and the latter extends to a significantly wider range in $x$ than the former. Magnetic ordering in this case is well removed from the collapse into the tcO phase. By comparing Figs. \ref{fig:comp_lattice_param}(a), \ref{fig:comp_lattice_param}(b) and \ref{fig:comp_lattice_param}(c), it becomes clear that the size of the $c$ lattice parameter by itself is not the factor for magnetism, since magnetic order appears: (a) for $11.42\ \text{\AA}<c<11.62\ \text{\AA}$ in Sr(Ni$_{1-x}$Co$_x$)$_2$P$_2$; (b) for $10.77\ \text{\AA}<c<11.33$ in SrCo$_2$(Ge$_{1-x}$P$_x$)$_2$; (c) for $9.59\ \text{\AA}<c<10.74\ \text{\AA}$ for Ca$_{1-x}$Sr$_x$Co$_2$P$_2$.

The case of Sr(Ni$_{1-x}$Co$_x$)$_2$P$_2$ shows that the emergence of magnetism need not coincide with the breaking or forming of the P-P bonds. In the case of SrCo$_2$(Ge$_{1-x}$P$_x$)$_2$, substitution of P with Ge causes the system to be electron deficient with respect to SrCo$_2$P$_2$, and in the case of Ca$_{1-x}$Sr$_x$Co$_2$P$_2$ the electron count remains the same since it corresponds to isovalent substitution. In contrast, substituting Co with Ni adds electrons to the system, which may be a relevant factor that enables Sr(Ni$_{1-x}$Co$_x$)$_2$P$_2$ to exhibit magnetism farther away from the lattice collapse, as opposed to the other two cases. In fact, itinerant ferromagnetism was also observed for electron doping in SrCo$_2$As$_2$ \cite{SHEN2019}.

\begin{figure}[H]
 \centering
 \includegraphics[width=\linewidth]{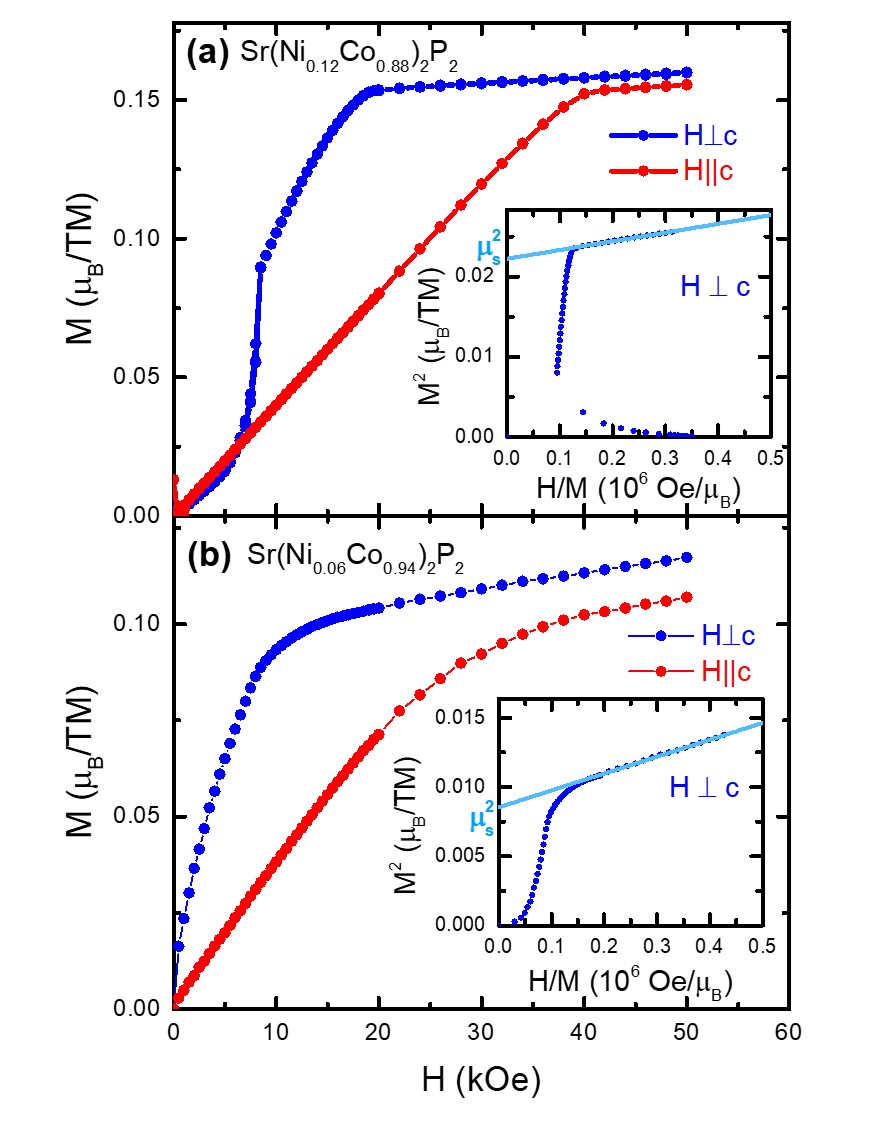}
 \caption{\footnotesize{(Main panels) Magnetization (in units of $\mu_B$ per transition metal) as a function of the applied magnetic field $H$ perpendicular (blue) and parallel (red) to the $c$ axis, measured at a constant temperature of $2\ \text{K}$, for (a) $x=0.88$ and (b) $x=0.94$. (Insets) $M^2$ as a function of $H/M$ for $H\perp c$; the light blue lines show the extrapolation of the high field behavior to zero field in order to estimate $\mu_s^2$.}}
 \label{fig:MH_2K_70n_example_2}
\end{figure}

In order to discuss the $M(H)$ isotherms shown in Fig. \ref{fig:MH_complete} in more detail, the data for $x=0.88$ and $x=0.94$ are plotted separately in the main panels of Fig. \ref{fig:MH_2K_70n_example_2} for both directions $H\perp c$ (blue) and $H||c$ (red). The data in Fig. \ref{fig:MH_2K_70n_example_2}(a) for $H\perp c$ (in blue) reveal the presence of metamagnetic transitions, which are present in the samples with $0.74 \leq x \leq 0.91$. This includes the jump in magnetization observed for $H \approx 8\ \text{kOe}$ as well as the change in slope for $H \approx 19\ \text{kOe}$, for $x=0.88$. The presence of metamagnetic transitions is a common characteristic displayed in systems with long-range finite $q$ magnetic order, and hence is consistent with the proposed AFM order.

\begin{figure}[H]
 \centering
 \includegraphics[width=\linewidth]{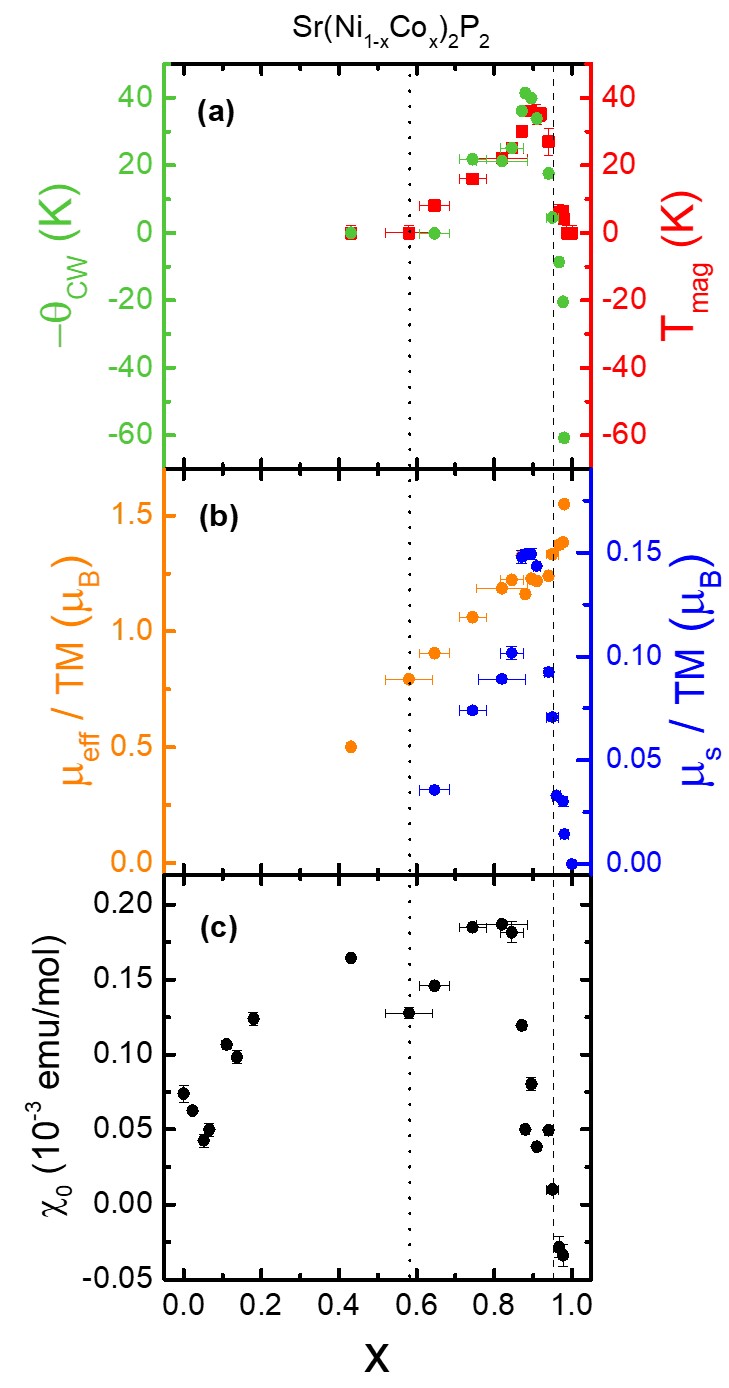}
 \caption{\footnotesize{(a) Curie-Weiss temperature $-\theta_{CW}$ (green circles, left axis), magnetic ordering temperature $T_{mag}$ (red squares, right axis); (b) effective moment $\mu_{\text{eff}}$ per transition metal atom (orange circles, left axis), $\mu_s$ per transition metal atom (blue circles, right axis); and (c) $\chi_0$ values, all as a function of the Co fraction $x$. The composition marked with a dotted line shows the emergence of the AFM order, and the dashed line shows the boundary between AFM and FM order.}}
 \label{fig:all_parameters}
\end{figure}

The values of the extrapolated moment, $\mu_{\text{s}}$, can be found from the $M(H)$ isotherms. The main panel of Fig. \ref{fig:MH_2K_70n_example_2}(b) reflects most clearly the fact that the high field behavior of $M$ vs $H$ is not strictly linear. Instead, as shown in the insets of Fig. \ref{fig:MH_2K_70n_example_2}, the high field data for $H\perp c$ follow a linear dependence of $M^2$ with $H/M$, as expected for an itinerant magnet at low temperatures \cite{takahashi1986origin}. For all compositions in the range $0.65\leq x\leq 0.98$, the square roots of the $y$-intercepts of the linear fits are taken as estimates for $\mu_s$, and its values are shown in Table \ref{table:Takahashi} and plotted in Fig. \ref{fig:all_parameters}(b). The value of $\mu_{\text{s}}$ increases from zero to a maximum value of $0.150\ \mu_B/TM$ for $x=0.88$, and then decreases back to zero as it approaches the composition of SrCo$_2$P$_2$. This behavior resembles that of the magnetic transition temperature with $x$ [Fig. \ref{fig:all_parameters}(a), red symbols].

In addition, Curie-Weiss dependence of some of the samples was examined in detail. Polycrystalline averages of the magnetic susceptibility $\chi$ were extracted from the high temperature dependence of the DC magnetic susceptibility $\chi=M/H$ at $H=10\ \text{kOe}$ for $H$ applied perperndicular ($\chi_{\perp}$) and parallel ($\chi_{||}$) to the $c$ axis. The temperature dependencies of $\chi$ are plotted in red (left axis) in Fig. \ref{fig:all_CW_fits} in Appendix B. Given that, for the AFM compounds, this field is comparable to the metamagnetic field for $H {\perp} c$ (e.g., see Fig. \ref{fig:MH_2K_70n_example_2}), the presented data can only be considered as $\chi(T)$ for temperatures well above $T_N$ or $T_C$. As such, we only analyze our data for $100\ \text{K}<T<300\ \text{K}$.
A nonlinear fit was applied for the polycrystalline averaged susceptibilities, according to the following the equation
\begin{equation}
    \chi=\frac{C}{T-\theta_{CW}}+\chi_0.
\end{equation}
The best fitted curve is plotted with a black line. Figure \ref{fig:all_CW_fits} also shows $(\chi-\chi_0)^{-1}$ in blue (right axis), for $0.58\leq x\leq 0.98$. The linear behavior of $(\chi-\chi_0)^{-1}$ for the fitting range supports the correspondence of the observed behavior of $\chi(T)$ to the fitted model. The temperature dependence measured for samples with $x=0.99$ and $x=1$ does not follow Curie-Weiss (CW) law, consistent with previous reports on pure SrCo$_2$P$_2$ \cite{Teruya2014}. 

The parameters obtained from our CW fits for $0.58<x<0.98$ are summarized in the Table \ref{table:allparameters_new}, as well as in Fig. \ref{fig:all_parameters}. The values of $\theta_{CW}$ and $\chi_0$ were obtained directly from the fit, whereas $\mu_{\text{eff}}$ was obtained from the fitted Curie constant $C$, according to
\begin{equation}
    C=\frac{N_A\mu_{\text{eff}}^2}{3k_B},
\end{equation}
where $N_A$ is the Avogadro number and $k_B$ is Boltzmann constant. The resulting $\mu_{\text{eff}}$ in units of emu per atom of transition metal was divided by $\mu_B=9.27\times 10^{-21}\ \text{emu}$ in order to express it in units of Bohr magnetons, $\mu_B$, per transition metal atom. The values of $\chi_0$ for $x<0.43$ were estimated as $\chi_0=\chi(300\ K)$, since those data did not fit to a Curie-Weiss behavior but were almost constant at high temperatures. This is not the case for $x\geq 0.99$ that do not follow Curie-Weiss law, but the relative variations in $\chi$ are large, hence $\chi_0$ was not estimated for those compositions. The error bars were obtained by propagating the standard errors of the fitting parameters. In the case of $\mu_{eff}$ and $\chi_0$ an additional contribution to the error was obtained by propagating the error in the mass of the measured crystals, whereas for $\theta_{CW}$ an error of $0.1\ \text{K}$ in the measurement of temperature was added to the standard error of the fit. The reason for the anomalously high value of $\theta_{CW}$ for the composition $x=0.58$ right below the emergence of magnetic ordering is unclear, although a similar behavior was observed in Sr(Co$_{1-x}$Ni$_x$)$_2$As$_2$ \cite{Sangeetha2019}.

\begin{table}[htbp]
  \centering
  \setlength{\tabcolsep}{3pt}
  \renewcommand{\arraystretch}{1.75}
    \begin{tabular}{|c| c c c c|}
    \hline
    $x$     &   $T_{\text{mag}}$   &  $\theta_{CW}$     & $\mu_{eff}/TM$ & $\chi_0 \times 10^6$\\
         &  $(K)$   &  $(K)$     & $(\mu_B)$ & $(emu/mol)$\\
    \hline
    0     &  0(2)   &  $-$     & $-$   &  74(6)\\
    0.023     &    0(2)   &  $-$     & $-$   &  $-$\\
    0.023     &    0(2)   &  $-$    & $-$  &  62.6(7)\\
    0.028     &    0(2)   &  $-$     & $-$   &   $-$\\
    0.052    &    0(2)   &  $-$     & $-$   &   43(4)\\
    0.066     &    0(2)   &  $-$     & $-$   &   50(4)\\
    0.11     &    0(2)   &  $-$     & $-$   &  107(3)\\
    0.14     &    0(2)   &  $-$     & $-$   &   98(4)\\
    0.18     &    0(2)   &  $-$    & $-$   &  124(4)\\
    0.43     &    0(2)   &  0.0(1)     & 0.470(7)   &   164(3)\\
    0.58     &    0(2)   &  102(4)     & 0.79(2)  &  128(3)\\
    0.65     &   8(1)   &  0.2(4)     &  0.897(5)  &   146(1)\\
    0.74     &    16(1)   &  -21.8(2)     & 1.053(8)  &  185(3)\\
    0.82     &    22(1)   &  -21.2(3)     & 1.177(8)  &  183(4)\\
    0.85     &    25(1)   &  -25.1(8)     & 1.21(2)  &  180(10)\\
    0.87     &    30(1)   &  -36.1(2)     & 1.31(1)  &  118(4)\\
    0.88     &    36(1)   &  -41.4(2)     & 1.151(8)  & 49(1)\\
    0.90     &    36(1) &  -39.8(4)     & 1.22(2)  &  78(6)\\
    0.91     &    35(3)   &  -33.8(2)     & 1.208(5)  &  38(1)\\
    0.94     &    27(4)   &  -17.6(3)     & 1.23(1)  &  48(3)\\
    0.95     &    $-$   &  -4.5(4)     & 1.322(2)  &  98(3)\\
    0.96     &    6.4(6)   &  9(1)     & 1.36(1)  & -28(7)\\
    0.98     &    6.6(4)   &  21(1)     & 1.37(6)  &  -34(10)\\
    0.98     &    4.3(7)   &  61(1)     & 1.54(1)  &  -127(8)\\
    0.99     &    0(2)   &  $-$     & $-$  &  $-$\\
    1     &    0(2)   &  $-$     & $-$  &  $-$\\
    
          \hline
    \end{tabular}%
    \caption{\footnotesize Parameters obtained from the nonlinear Curie-Weiss fits, and magnetic transition temperatures. For $x<0.43$ and $x>0.98$ the data did not correspond to the fitting model, so the fitting parameters are not included in the table. The value of $\chi_0$ for $x<0.43$ was estimated as $\chi_0=\chi(300\ K)$. As mentioned in the text, the two compositions listed as $x=0.98$ correspond to different batches and, despite not being able to resolve their differences by EDS, they show differences in their properties.}
  \label{table:allparameters_new}%
\end{table}%

As shown in Fig. \ref{fig:all_parameters}, values of $\mu_{\text{eff}}$ increase as a function of $x$, whereas the values of $-\theta_{CW}$ follow the nonmonotonic behaviour of $T_N$ and, therefore, its dependence is reminiscent of the one manifested in the phase diagram (Fig. \ref{fig:phase_diagram}). Moreover, the values of $\mu_{\text{eff}}$ are significantly larger that the values of $\mu_{s}$ shown in Fig. \ref{fig:all_parameters}(b). A local maximum of $\chi_0$, which can ultimately be related to the density of states near the Fermi level, occurs at similar compositions as the maximum of $T_{mag}$, $-\theta_{CW}$ and $\mu_s$. 

Given that values of $\mu_s$ in these metallic samples are small compared to $\mu_{eff}$, it should not be surprising that the observed behavior is in good agreement with the spin fluctuation theory of itinerant electron magnetism \cite{Takahashi}, which, though in principle, only applicable to weak itinerant ferromagnets, has been extended to itinerant antiferromagnets as well \cite{Sangeetha2019}. Therefore, it is reasonable to rationalize the effect of Co substitution on the magnetic properties in terms of band filling, rather than changes in the exchange couplings of an anisotropic Heisenberg model as argued for local moment systems.

\begin{table}[htbp]
  \centering
  \setlength{\tabcolsep}{1.5pt}
  \renewcommand{\arraystretch}{1.75}
    \begin{tabular}{|c|c c c|c c c|}
    \hline
    $x$     &    $T_C$   &  $T_0$  & $T_C/T_0$   & $\mu_{eff}$  & $\mu_{s}$ & $\mu_{eff}/\mu_{s}$\\
         &    (K)   &  (K)  &    & $(\frac{\mu_B}{TM})$  & $(\frac{\mu_B}{TM})$ & \\
    \hline
    0.65     &   5.6(2)*  & 390(10)  &  0.014(3)*  &  0.897  &  0.0357(4)  & 25.3(4)\\
    0.74     &    $-$   &  $-$  & $-$  & 1.053  & 0.074(1) & 14.3(3)\\
    0.82     &    $-$   &  $-$  & $-$  & 1.177  & 0.089(1) & 13.3(3)\\
    0.85     &    21(4)*   &  470(30) &  0.045(5)*
    & 1.21  & 0.101(3) & 12.1(6)\\
    0.87     &    29(1)*   &  380(20) &  0.076(7)*  & 1.31  & 0.148(3) & 9.0(3)\\
    0.88     &    $-$   &  $-$  & $-$  & 1.151  & 0.150(1) & 7.8(2)\\
    0.90     &    $-$ &  $-$ & $-$ & 1.22  & 0.150(4) & 8.2(3)\\
    0.91     &    $-$   &  $-$ &  $-$  & 1.208  & 0.144(1) & 8.5(1)\\
    0.94     &    $-$   &  $-$ &  $-$  & 1.23  & 0.092(1) & 13.0(3)\\
    0.95     &    $-$   &  $-$ & $-$   & 1.322  & 0.071(1) & 18.9(5)\\
    0.96     &    6.4(6)   &  350(10) &  0.019(4)  & 1.36  & 0.0501(9) & 27.4(7)\\
    0.98     &    6.6(4)   &  800(100) &  0.008(3)  & 1.37  & 0.030(3) & 46(5)\\
    0.98     &    4.3(7)   &  1330(80) &  0.0032(9)  & 1.54  & 0.0145(3) & 107(3)\\

          \hline
    \end{tabular}%
    \caption{\footnotesize Parameters used to build the Takahashi-Deguchi plot of figure \ref{fig:Takahashi_FM}. The value marked with * corresponds to $T_C^*$ estimated for the metamagnetic phase at high fields, explained in detail in Appendix C.}
  \label{table:Takahashi}%
\end{table}%

\begin{figure}[H]
 \centering
 \includegraphics[width=\linewidth]{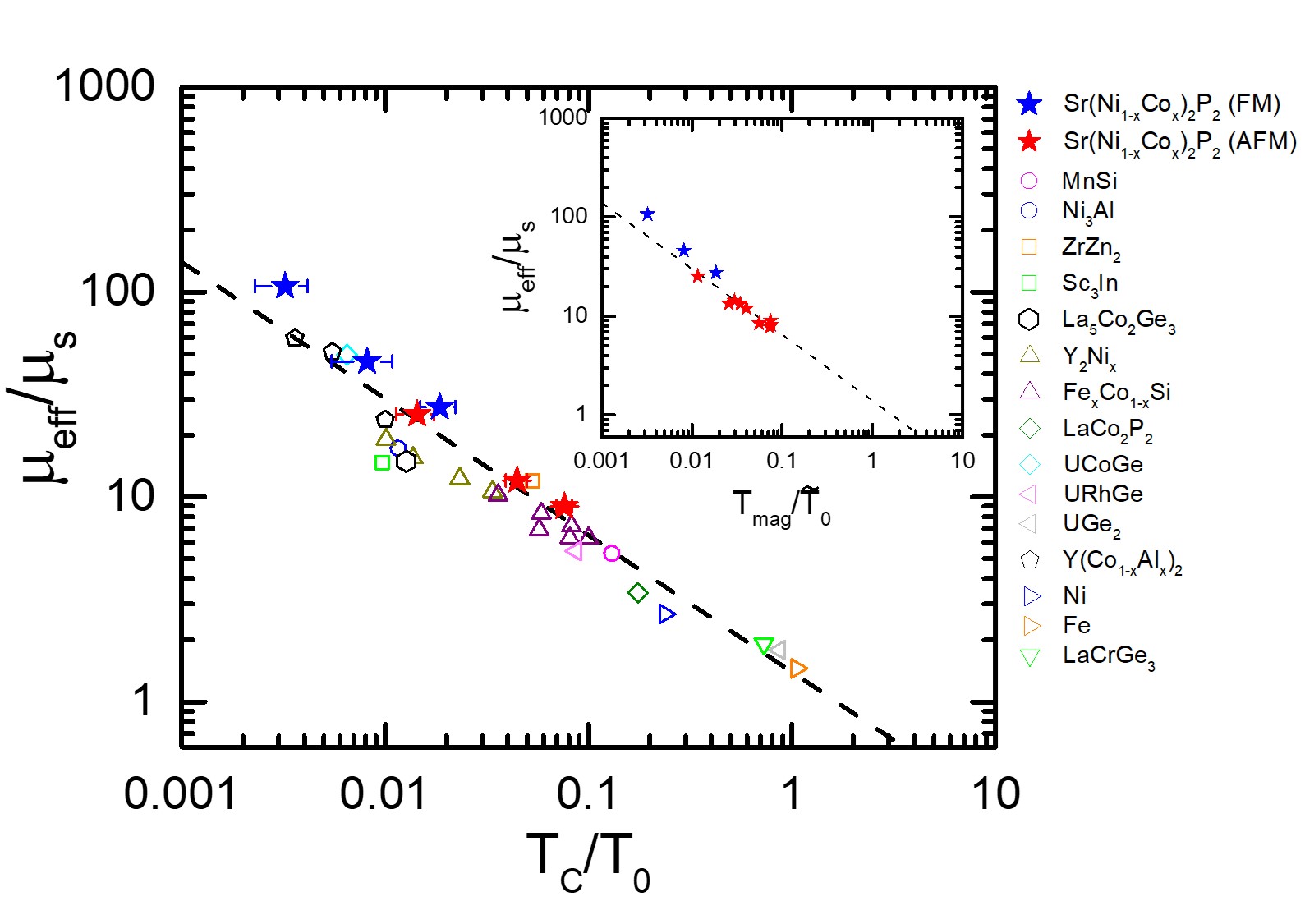}
 \caption{\footnotesize{(Main panel) Deguchi-Takahashi plot for the three FM compositions (solid blue stars) and the AFM compositions $x=0.65$, $x=0.85$ and $x=0.87$ (solid red stars), as well as other itinerant magnets taken from the references \citenum{takahashi1986origin}, \citenum{Takahashi}, \citenum{Saunders2020} and \citenum{Xu2023}. The dashed line represents the theoretically expected dependence given by equation \ref{eq:theory}. (Inset) Modified Deguchi-Takahashi plot, for the FM (solid blue stars) as well as AFM (solid red stars) samples.}}
 \label{fig:Takahashi_FM}
\end{figure}

In order to show this good agreement, a Deguchi-Takahashi plot is presented in the main panel of Fig. \ref{fig:Takahashi_FM}, where $T_C$ is the FM Curie temperature, and $T_0$ is a spectral parameter corresponding to the width of the frequency dependence of the generalized susceptibility at the zone boundary $q_B$ or, in other words, to the lifetime of spin fluctuations with that $q$. The values of $T_0$ are included in Table \ref{table:Takahashi}, which have been calculated from the slope ($s$) and intercept ($\mu_s^2$) of the Arrott plot at $2\ \text{K}$, according to
\begin{equation}
     T_0=\left(\frac{20T_C}{\mu_s^2}\right)^{6/5}\left(248\frac{\text{Oe}}{\text{K}}\ s\right)^{3/5},
    \label{eq:T0}
\end{equation}
 explained more detail in Appendix C. The three compositions of Sr(Ni$_{1-x}$Co$_x$)$_2$P$_2$ identified as FM were included in the plot in solid blue stars, together with other well-known itinerant ferromagnets \cite{takahashi1986origin,Takahashi,Saunders2020,Xu2023}. The main panel of the figure shows that the mentioned compositions not only belong to the same trend as many other itinerant systems, but also follow the theoretical expected behavior \cite{Takahashi}, represented by the dashed line, given by
 \begin{equation}
     \frac{\mu_{\text{eff}}}{\mu_s} \simeq 1.4 \left( \frac{T_C}{T_0} \right)^{2/3}.
     \label{eq:theory}
 \end{equation}

This theory has also been applied to systems that are not strictly ferromagnetic but, with a finite magnetic field applied, exhibit typical characteristics of weak itinerant ferromagnets \cite{takahashi1985, Sangeetha2019}. For the present work, the cases with $x=0.65$, $x=0.85$ and $x=0.87$ were analyzed in detail by constructing Arrott plots (see Figs. \ref{fig:arrot_DR831}-\ref{fig:arrot_DR821} in Appendix C). The criteria for determining $T_C^*$ of the high field state was applied by fitting the high field results, obtaining slightly lower values than the AFM ordering temperatures. The slope and intercept of the high field results at $T=2\ \text{K}$ were used to estimate the spectral parameter $T_0$, similar to what was done for the ferromagnetic samples. The results obtained (solid red stars in main panel of Fig. \ref{fig:Takahashi_FM}) also lay within the theoretical prediction represented by the dashed line, indicating that the magnetic ordering in this system is itinerant not only for $0.96<x<0.98$, but for lower compositions as well. 

Table \ref{table:Takahashi} also shows that $\mu_{\text{eff}}/\mu_s$ covers wide range for the system Sr(Ni$_{1-x}$Co$_x$)$_2$P$_2$, spanning about an order of magnitude. This occurs in a nonmonotonic way, with the values of $\mu_{\text{eff}}/\mu_{\text{s}}$ decreasing with $x$ until reaching a minimum for $x=0.88$ (near the maximum of the dome in the phase diagram), and then increasing until reaching its maximum of 107 for $x=0.98$. The values of $T_0$ are significantly greater than the ordering temperatures, consistent with weak itinerant magnetism for which this theory is applicable.

Assuming that $T_N$ and the Arrott-plot-inferred $T_C^*$ are close for the other AFM samples as well, we can plot all of our data on a modified Deguchi-Takahashi plot (see inset of Fig. \ref{fig:Takahashi_FM}) using $T_{mag}/\tilde{T}_0$ data, where $T_{mag}$ is taken as $T_C$ or $T_N$ for the FM and AFM samples, respectively. Since the calculation of $T_0$ requires the value of the Curie temperature as seen in Eq. \ref{eq:T0}, which was not estimated for all of the AFM samples, a new parameter was defined as
\begin{equation}
    \tilde{T}_0=\left(\frac{20T_{mag}}{\mu_s^2}\right)^{6/5}\left(248\frac{\text{Oe}}{\text{K}}\ s\right)^{3/5},
\end{equation}
The values of this parameter are included in Table \ref{tab:tilde} in Appendix C, which show a notably large dispersion since it relies on the assumed proximity between $T_N$ and $T_C^*$. However, the purpose of this analysis is to show the negative correlation between $\mu_{\text{eff}}/\mu_{s}$ and $T_{mag}/\tilde{T}_0$. The extent to which these data agree with the theoretical prediction represented with the dashed line is strongly dependent on the degree of similarity between $T_N$ and the inferred $T_C^*$ for the AFM samples, which was only evaluated for three of the AFM samples in this work.

\section{Conclusion}
\label{sec:Conclusion}

This study shows how Co substitution in Sr(Ni$_{1-x}$Co$_x$)$_2$P$_2$ can serve to gradually decrease the ucT $\leftrightarrow$ tcO transition, until suppressing it fully at $x\sim 0.1$. Further substitution ($0.11 \leq x \leq 0.58$) leads to having the ucT phase in the full temperature range explored, with no other phase transition detected for $2\ K\leq T \leq 300\ \text{K}$. For $0.64\leq x \leq 0.94$ antiferromagnetic ordering appears, initially with $T_N$ increasing with $x$, and ultimately decreasing and giving rise to ferromagnetic order for $0.96 \leq x\leq 0.98$. Curie temperature decreases monotonically with $x$ until full suppression, giving paramagnetic behavior for $x\geq 0.99$. 

By comparing the Sr(Ni$_{1-x}$Co$_x$)$_2$P$_2$, Ca$_{1-x}$Sr$_x$Co$_2$P$_2$ and SrCo$_2$(Ge$_{1-x}$P$_x$)$_2$ systems we find that there is no "special" $c$-axis value for the stabilization of the magnetic region. In addition, compared to Ca$_{1-x}$Sr$_x$Co$_2$P$_2$ and SrCo$_2$(Ge$_{1-x}$P$_x$)$_2$ systems, Sr(Ni$_{1-x}$Co$_x$)$_2$P$_2$ is the only one for which the region with AFM and/or FM ordering evolves well removed from any collapse of the structure. Magnetism in these systems appears to be well parametrized by Takahasi's theory for itinerant electron magnetism, as shown by the six compositions in which the Deguchi-Takahashi analysis was performed.

\begin{acknowledgements}
The authors acknowledge Mingyu Xu and John Titus William Barnett for discussion and assistance in some growths. Work was done at Ames National Laboratory was supported by the U.S. Department of Energy, Office of Basic Energy Science, Division of Materials Sciences and Engineering. Ames National Laboratory is operated for the U.S. Department of Energy by Iowa State University under Contract No. DE-AC02-07CH11358.

\end{acknowledgements}

\section*{Appendix A}

\begin{figure}[H]
 \centering
 \includegraphics[width=0.9\linewidth]{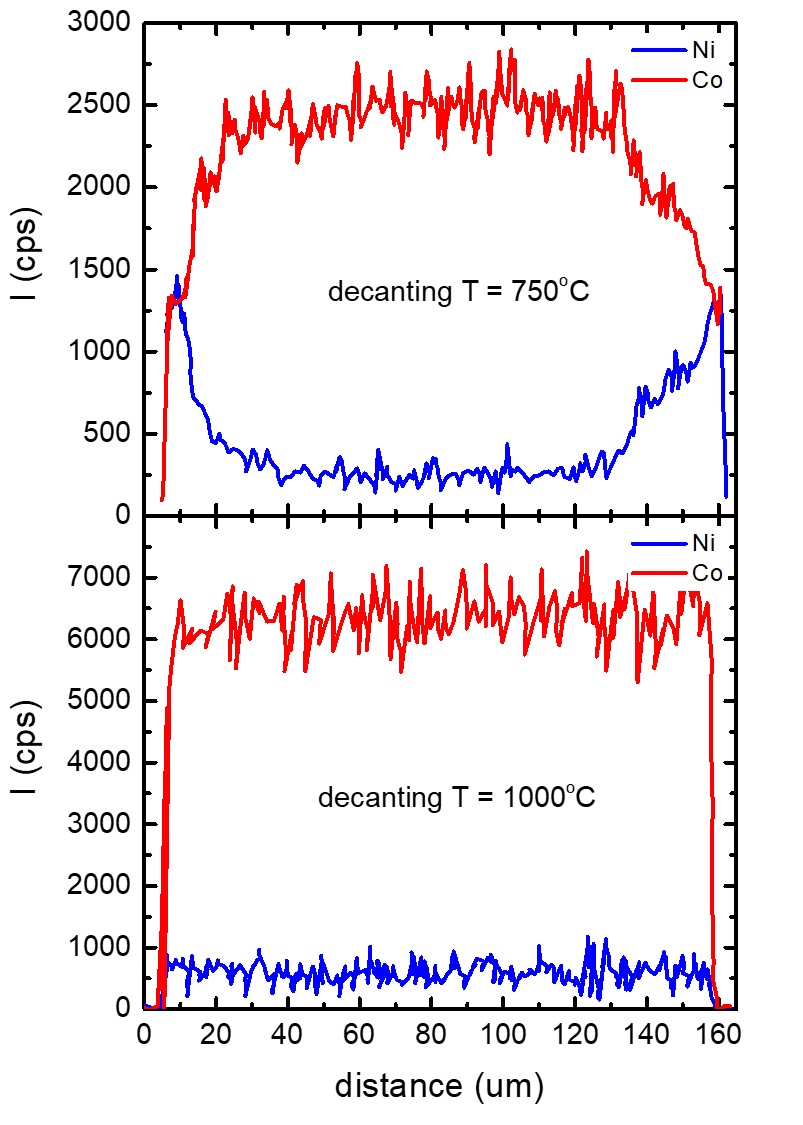}
 \caption{\footnotesize{ X-ray intensity, in counts per second, obtained from EDS measurement at the energies corresponding to Ni K-$\alpha$ (blue line) and Co K-$\alpha$ (red line), as a function of the distance along $(0 0 1)$ on a polished edge of a sample from the batch decanted at $750^{\circ}\text{C}$ (upper pannel) and a sample from the batch decanted at $1000^{\circ}\text{C}$ (lower pannel), both with the same nominal composition of Sr(Ni$_{0.4}$Co$_{0.6}$)$_2$P$_2$.}}
 \label{fig:inhomog}
\end{figure}

During the crystal growth process, when a low decanting temperature ($T \sim 700^{\circ}\text{C}$) was used, EDS measurements indicated inhomogeneous Co and Ni distributions across a polished edge of the crystal (along $c$). This is represented in the upper panel of Fig. \ref{fig:inhomog} by the variation of the x-ray intensity at the Ni (blue) and Co (red) $K\alpha$ lines. This suggests that the crystals grow richer in Co initially, and poorer as the growth develops further, which is consistent with the samples having higher Co fraction in the center, and lower towards the outer surfaces. A possible reason for this is the following: since the crystals initially tend to grow with a higher Co fraction that the nominal fraction contained in the crucible, as the furnace is cooled and the  crystals grow larger, the coexisting melt gets more depleted in Co, which leads to a lower Co fraction for the outer layers that grow later. Examining Fig. \ref{fig:inhomog}(a) suggested to us, then, that if we only cooled to $1000^{\circ}$C, we could effectively not grow the two edge regions where the large changes in the Ni-to-Co ratio took place. Indeed, based on this hypothesis, we managed to minimize the degree of inhomogeneity by decanting the flux at higher temperatures ($T\sim 1000^{\circ}\text{C}$), leading to fewer and/or smaller crystals, but with more homogeneous Co and Ni profiles as a function of depth, as the one shown in the lower panel of Fig. \ref{fig:inhomog}. Both examples provided correspond to the same nominal composition Sr(Ni$_{0.4}$Co$_{0.6}$)$_2$P$_2$.

There are two points that are important to note about this tendency toward inhomogeneity.  First, it is associated with growing the psuedoternary, quaternary compound out of a vast abundance of a fifth element (Sn) flux.  Given that there is a very limited amount of Ni and Co in the melt, as crystals grow they can rapidly deplete one element over another. This is a very different case from growing, for example, Ba(Fe$_{1-x}$Co$_x$)$_2$As$_2$ out of a quaternary melt where there is a vast excess of (Fe$_{1-x}$Co$_x$)As melt [e.g., out of a Ba$_x$(Fe/Co)$_y$As$_z$ melt]. In this case the growth of a small amount of crystal phase with a $x_{EDS}$ different from $x_{\text{nominal}}$ does not shift the melt stoichiometry as much. Second, the changing relative amount of Ni and Co along the width of the platelike crystal shown in Fig. \ref{fig:inhomog}(a) suggests that this could be a method to create samples with tailored composition profiles if needed.

\section*{Appendix B}

\begin{figure}[H]
 \centering
 \includegraphics[width=\linewidth]{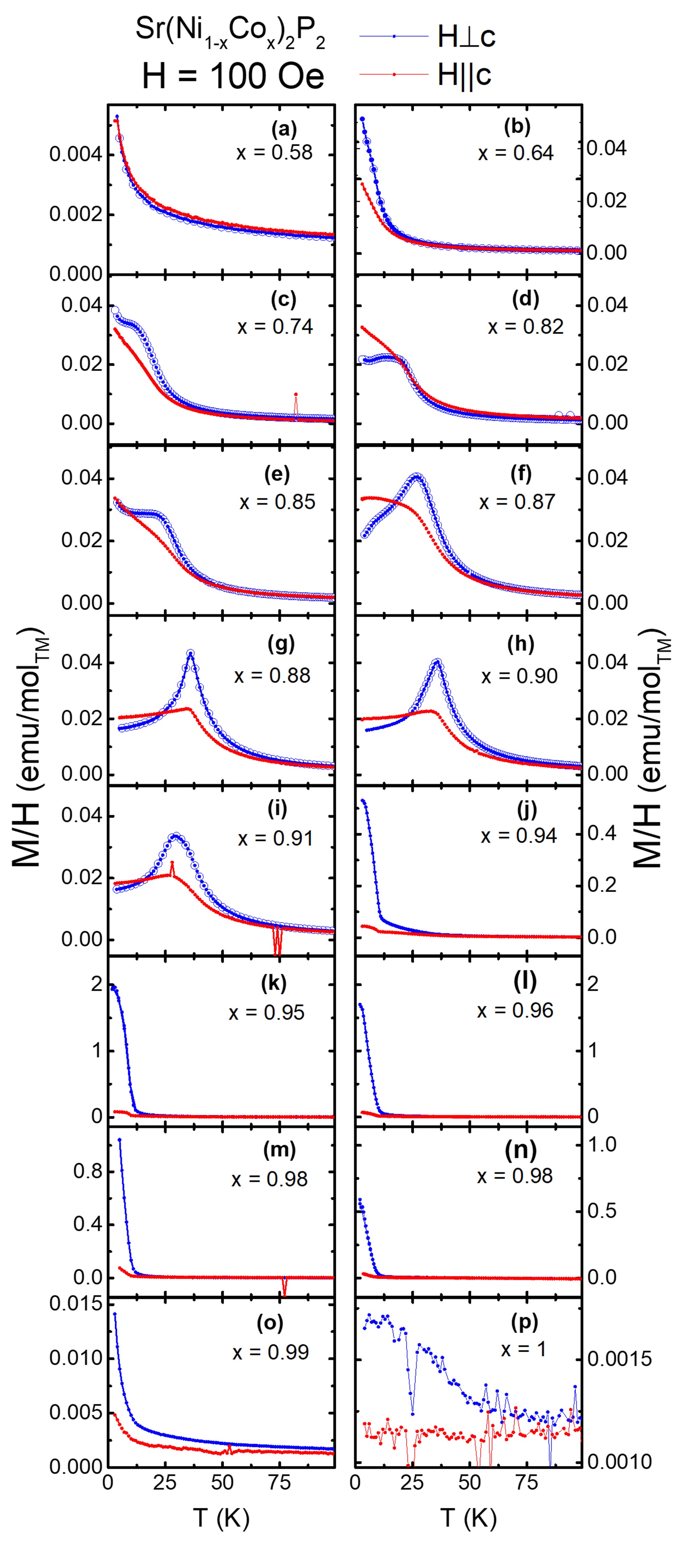}
 \caption{\footnotesize{Temperature dependent DC magnetization divided by applied field for compositions with different Co fraction $x$, measured with an applied field of $100\ \text{Oe}$ perpendicular (blue) and parallel (red) to the $c$-axis. For the samples with $0.58\leq x\leq 0.91$ with $H\perp c$, both ZFC (closed symbols) and FC (bigger open symbols) measurements were performed. For the rest, only ZFC measurements (closed symbols) were done.}}
 \label{fig:MT_100Oe_alltogether}
\end{figure}

The emergence of magnetic ordering in Sr(Ni$_{1-x}$Co$_x$)$_2$P$_2$ for $x>0.58$ can be further appreciated in Fig. \ref{fig:MT_100Oe_alltogether}. This figure shows the temperature-dependent magnetic DC susceptibility for $0.58 \leq x \leq 0.99$ samples, measured with a field of $100\ \text{Oe}$ applied perpendicular (in blue) as well as parallel (in red) to the $c$ axis. The results show varied magnetic behavior of the samples, ranging from paramagnetic (e.g. $x=0.58$) to antiferromagnetic (e.g., $x=0.88$) to ferromagnetic (e.g., $x=0.96$). 

\begin{figure}[H]
 \centering
 \includegraphics[width=0.9\linewidth]{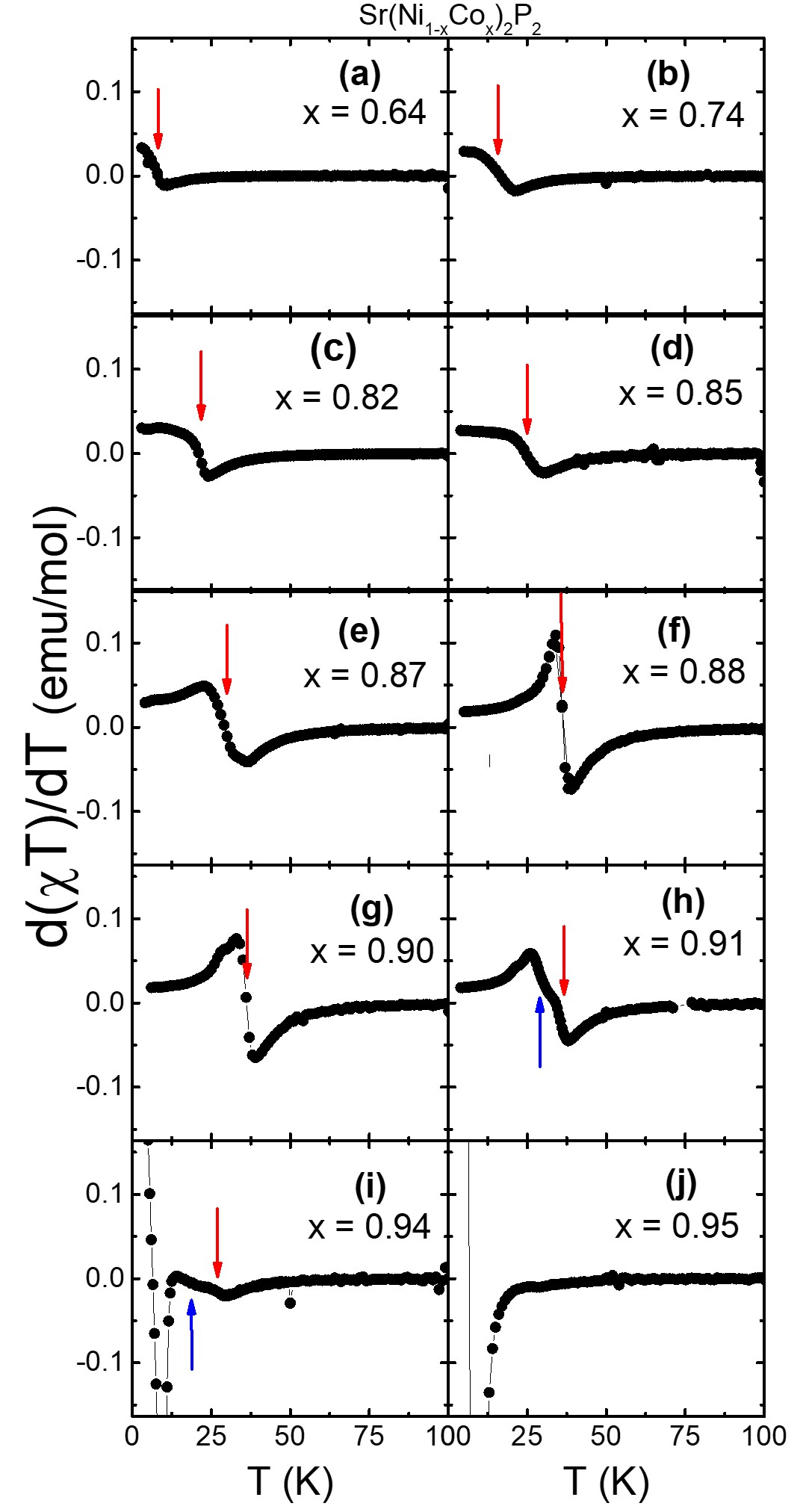}
 \caption{\footnotesize{Temperature derivative of $\chi_{\text{ave}} T$ for all the compositions identified as antiferromagnetic in the current work. The AFM transition temperatures are indicated with vertical arrows.}}
 \label{fig:dchiTdT_all}
\end{figure}

The criteria used for determining $T_N$ are shown in Fig. \ref{fig:dchiTdT_all} for all the antiferromagentic samples studied in this work. In the same way as was presented in the inset of Fig. \ref{fig:MT_RT_70n_example}(a) for $x=0.88$, the figure includes plots of $d\chi T/dT$ as a function of temperature for the rest of the compositions in which the criteria were used. $T_N$ was determined as the temperature at which the slope of  the function $f(T)=d(\chi_{\text{ave}} T)/dT$ is maximum. More specifically, the point for which $(f(T_i)-f(T_{i-1}))/(T_i-T_{i-1})$ was largest, and $T_i-T_{i-1}$ was taken as the uncertainty. For those compositions in which $(f(T_i)-f(T_{i-1}))/(T_i-T_{i-1})$ was similar for a temperature range larger than $T_i-T_{i-1}$, that whole range was taken as the uncertainty. It additionally includes the case of $x=0.95$, for which the steplike feature is suppressed (or at least no longer detected). The transition temperatures are indicated with a vertical arrows. The compositions $x=0.91$ and $x=0.94$ apparently present two steps separated by $\sim 5\ \text{K}$ from each other, so two arrows of different colors were drawn for those cases. The underlying reason for observing this is not clear: it could be due to the inhomogeneity of the Co concentration in the sample, or due to the intrinsic occurrence of two transitions. Further studies are needed in order to answer this. For this work, in those cases only one point was plotted in the phase diagram in Fig. \ref{fig:phase_diagram}, taken as the average between the two arrows, and the uncertainty such that it includes both arrows.

\begin{figure}[H]
 \centering
 \includegraphics[width=\linewidth]{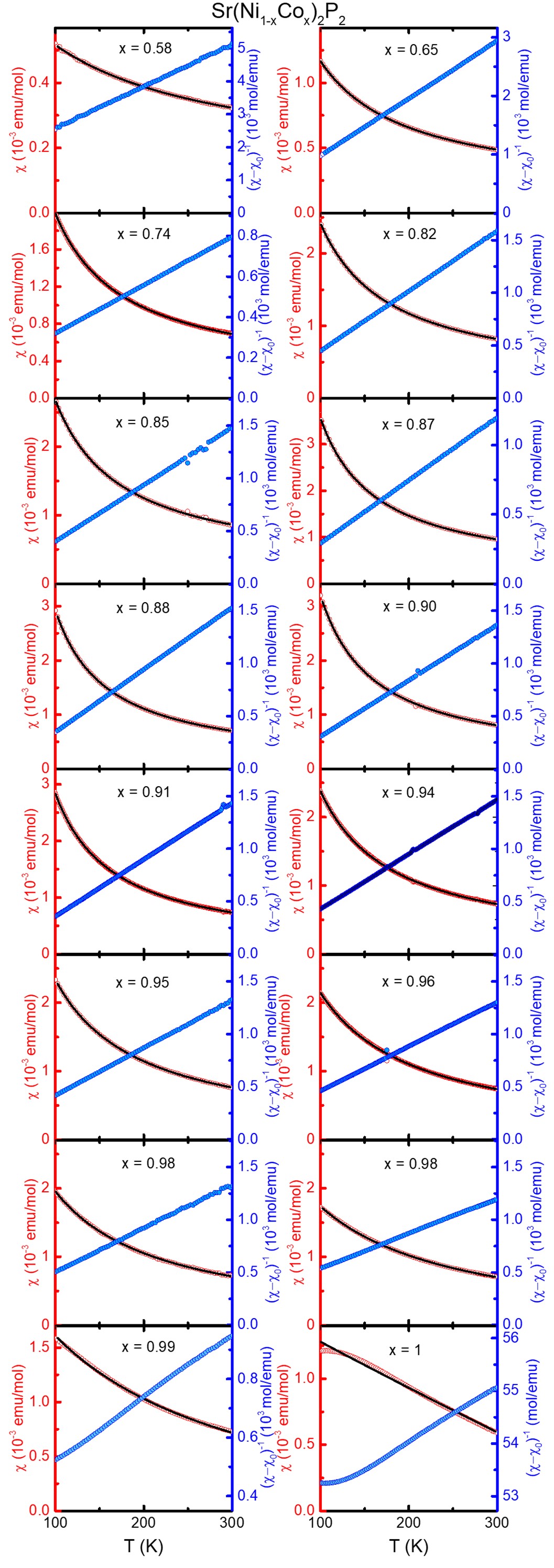}
 \caption{\footnotesize{Polycrystalline average of the DC magnetic susceptibility $\chi$ (red, left axis) as a function of temperature, measured with an applied field of $10\ \text{kOe}$, and the Curie-Weiss fit with a solid line (black, left axis). The values of $(\chi-\chi_0)^{-1}$ as a function of temperature are plotted in blue (right axis). As mentioned in the text, the two compositions listed as $x=0.98$ correspond to different batches and, despite not being able to resolve their differences by EDS, they show differences in their properties.}}
 \label{fig:all_CW_fits}
\end{figure}

The temperature dependencies of $\chi_{ave}$ are plotted in red (left axis) in Fig. \ref{fig:all_CW_fits}, together with $(\chi-\chi_0)^{-1}$ in blue (right axis), for $0.58\leq x\leq 0.98$, for an applied field of 10 kOe in each direction, for $100\ \text{K}<T<300\ \text{K}$.
A nonlinear fit was applied for the polycrystalline averaged susceptibilities, according to the following the equation
\begin{equation}
    \chi_{ave}=\frac{C}{T-\theta_{CW}}+\chi_0.
\end{equation}
The best fitted curve is plotted with a black line. The correspondence of the observed behavior to the fitted model can be further supported by noticing the linear behavior of $(\chi-\chi_0)^{-1}$ for the fitting range. The temperature dependence measured for samples with $x=0.99$ and $x=1$ does not follow Curie-Weiss law in the range $100\leq T\leq 300\ K$, as manifested by the difference between the fit (black line) and $\chi(T)$ (open red circles), as well as the nonlinearity of the estimated $(\chi-\chi_0)^{-1}$. These indications are subtle for $x=0.99$ and more obvious for $x=1$.

\section*{Appendix C}

\begin{figure}[H]
 \centering
 \includegraphics[width=\linewidth]{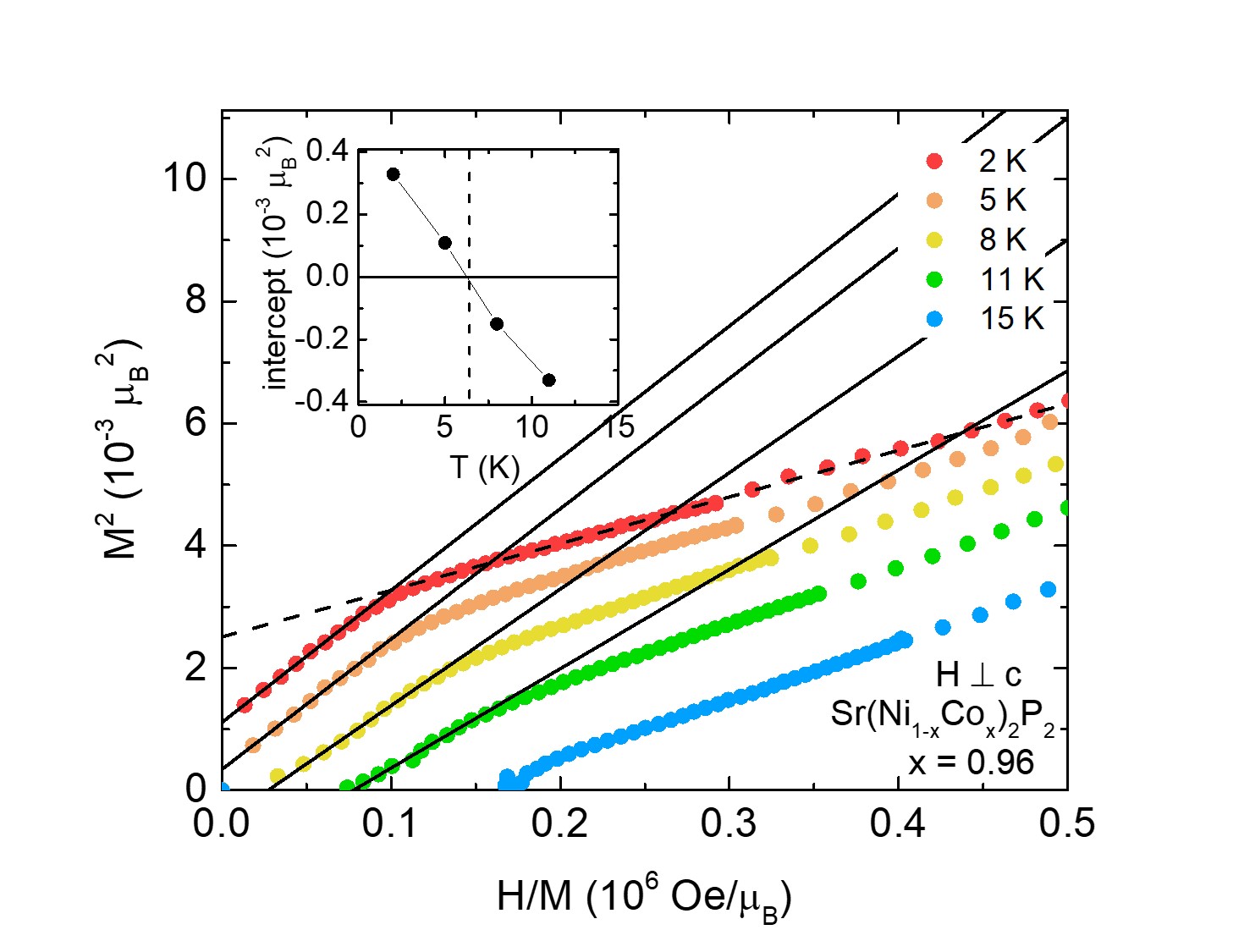}
 \caption{\footnotesize{Main panel: $M^2$ as a function of $H/M$ (Arrott plot) at different constant temperatures for a sample with $x=0.96$, and their corresponding linear fits (solid lines). Inset: Value of the intercept obtained from the linear fits for the different temperatures. The dashed line corresponds to the fit used to determine $\mu_s$ and $T_0$.}}
 \label{fig:arrot_PR598}
\end{figure}

\begin{figure}[H]
 \centering
 \includegraphics[width=\linewidth]{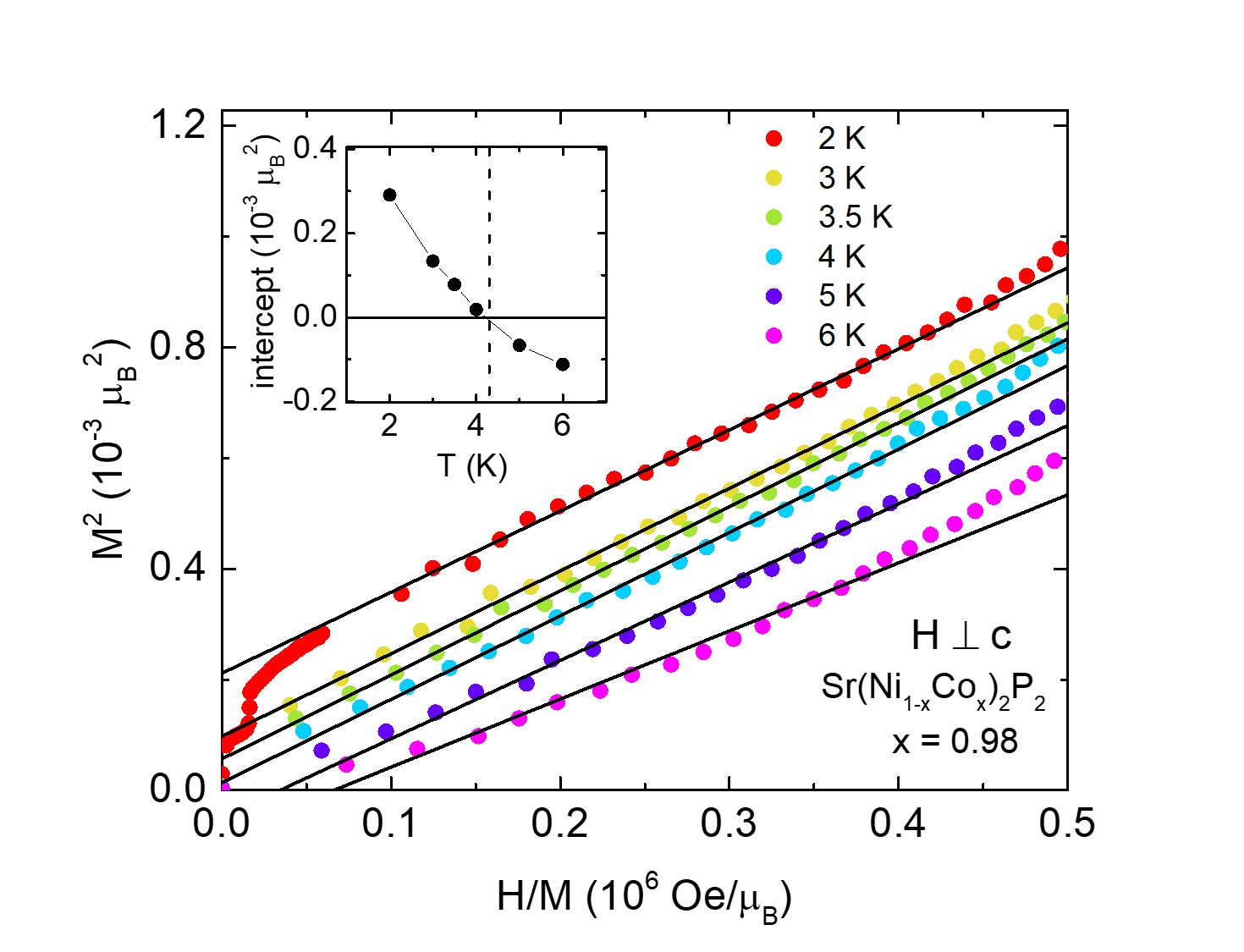}
 \caption{\footnotesize{Main panel: $M^2$ as a function of $H/M$ (Arrott plot) at different constant temperatures for a sample with $x=0.98$, and their corresponding linear fits (solid lines). The presence of a small step-like feature at $H\sim 200\ \text{Oe}$ is due to the superconducting signal of Sn impurity, and is emphasized when plotting $M^2$ in an Arrott plot. Inset: Value of the intercept obtained from the linear fits for the different temperatures.}}
 \label{fig:arrot_TM174}
\end{figure}

Arrott plots were constructed in order to determine the $T_C$ of the three compositions identified as ferromagnetic in this work. The results are plotted in Figs. \ref{fig:arrot_PR598} and \ref{fig:arrot_TM174}, in a similar way as done in Fig. \ref{fig:Arrot} in the main text. 

\begin{figure} [H]
 \centering
 \includegraphics[width=\linewidth]{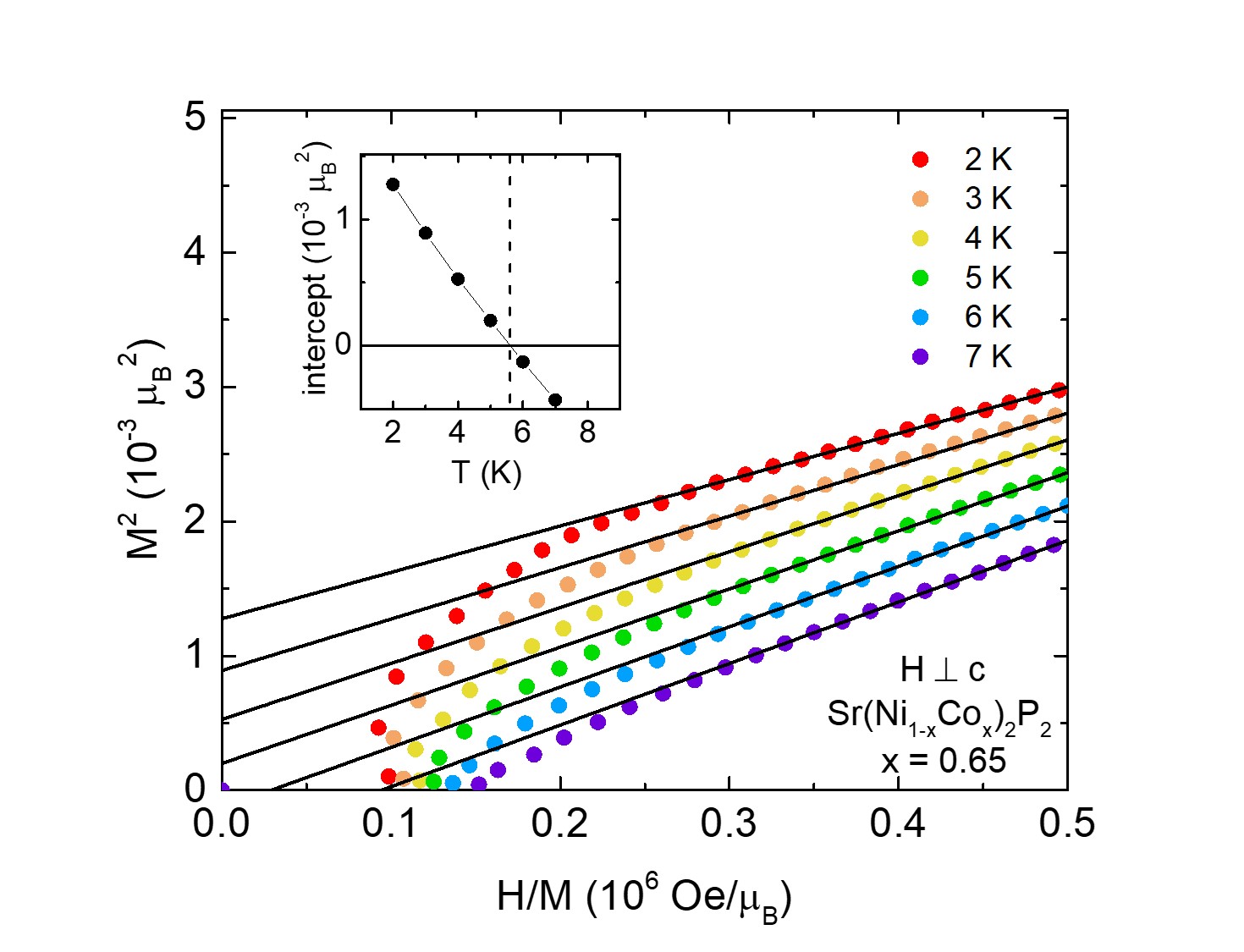}
 \caption{\footnotesize{Main panel: $M^2$ as a function of $H/M$ (Arrott plot) at different constant temperatures for a sample with $x=0.65$, and their corresponding linear fits (solid lines). Inset: Value of the intercept obtained from the linear fits for the different temperatures.}}
 \label{fig:arrot_DR831}
\end{figure}

\begin{figure} [H]
 \centering
 \includegraphics[width=\linewidth]{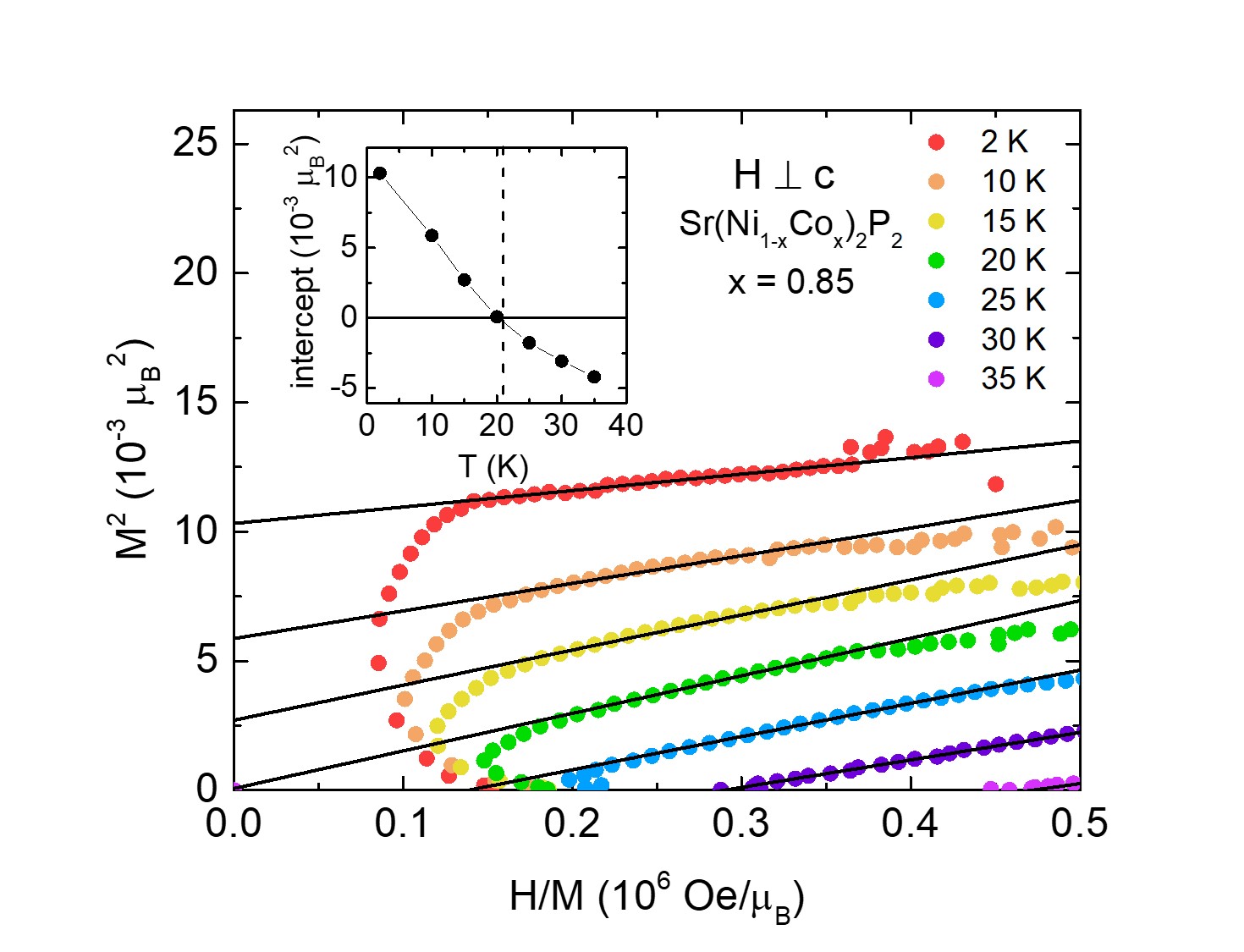}
 \caption{\footnotesize{Main panel: $M^2$ as a function of $H/M$ (Arrott plot) at different constant temperatures for a sample with $x=0.85$, and their corresponding linear fits (solid lines). Inset: Value of the intercept obtained from the linear fits for the different temperatures.}}
 \label{fig:arrot_DR820}
\end{figure}

\begin{figure} [H]
 \centering
 \includegraphics[width=\linewidth]{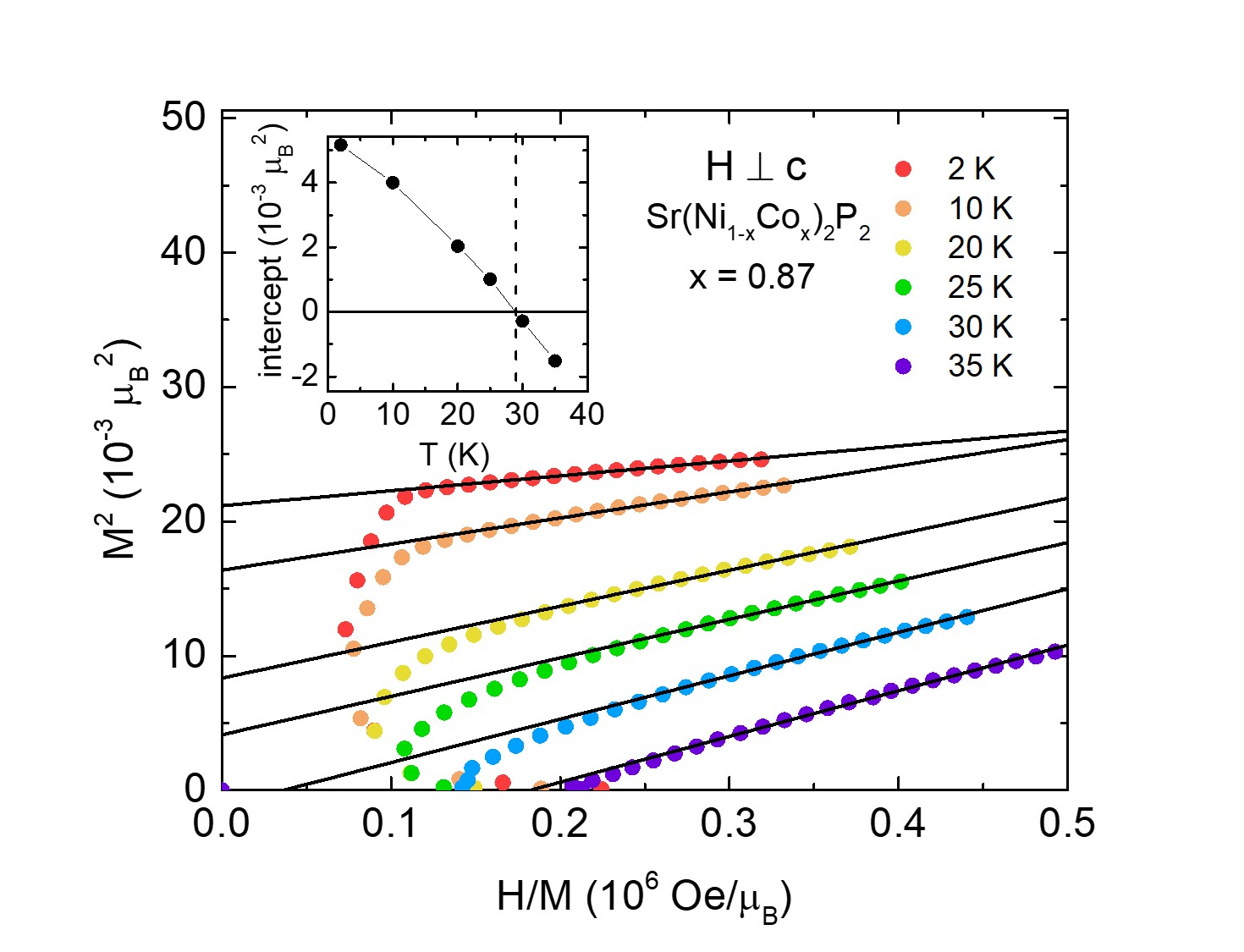}
 \caption{\footnotesize{Main panel: $M^2$ as a function of $H/M$ (Arrott plot) at different constant temperatures for a sample with $x=0.87$, and their corresponding linear fits (solid lines). Inset: Value of the intercept obtained from the linear fits for the different temperatures.}}
 \label{fig:arrot_DR821}
\end{figure}

Figure \ref{fig:arrot_PR598} presents a change in slope of $M^2$ vs $H/M$ at low fields. This feature is intrinsic to the behavior of the sample, and it shares some similarity to the changes in slope of Figs. \ref{fig:arrot_DR831}-\ref{fig:arrot_DR821} that occur due to the metamagnetic transitions in the AFM samples. It should be noted that, despite the fact that the samples with $x=0.95$ and $x=0.96$ show a clear FM component as shown by the orange and yellow curves in Fig. \ref{fig:MH_complete}(c), they may exhibit a more complex type of ordering, as their compositions are near the boundary between AFM and FM. Nevertheless, by plotting $M^2$ as a function of $H/M$, the field-dependent behavior below this feature as well as the behavior above that feature are successfully linearized, as expected for a mean-field theory. For the purpose of determining the Curie temperature, $T_C$, linear fits (solid lines) were performed for fields below this change of slope.

Figures \ref{fig:arrot_DR831}-\ref{fig:arrot_DR821} are modified Arrott plots for $x=0.65$, $x = 0.85$ and $x=0.87$, respectively, which actually order antiferromagnetically in a small applied field, but their behavior above the metamagnetic transition resembles that of the other ferromagnetic samples. Therefore, linear fits were performed for the higher field data, and used for the Deguchi-Takahashi analysis. This plot provides a way of estimating a value of $T_C^*$ for the field-induced state.
%as well as estimating the value of $T_0$. 

The Arrott plots can be used in order to determine the spectral parameter $T_0$. According to SCR (self-consistent renormalization) spin fluctuation theory \cite{Takahashi}, magnetic isotherms for itinerant ferromagnets follow the equation
\begin{equation}
    (g\mu_B)H=F_1\left[\frac{M^2}{(N_0g\mu_B)^2}- \frac{\mu_s^2}{(N_0g\mu_B)^2}\right] \frac{M}{N_0g\mu_B}
\end{equation}
where $F_1$ is the quartic coefficient in the Guizburg-Landau expansion of the free energy, $g$ is the Land\'e factor, $N_0$ the number of moment-bearing ions, and $\mu_s$ is the spontaneous magnetization. This expression can be rearranged to
\begin{equation}
    \left( \frac{M}{N_0\mu_B} \right)^2=\left( \frac{\mu_s}{N_0\mu_B} \right)^2+\frac{g^4 \mu_B}{F_1}\frac{H}{\left(\frac{M}{N_0\mu_B}\right)},
\end{equation}
so that it explicitly reflects the linear behavior of an Arrott plot. If $M$ and $\mu_s$ are written in units of $\mu_B$ per magnetic ion, and $H$ in Oe, then the slope ($s$) and intercept ($\mu_s^2$) of the Arrott plot can be written as 
\begin{equation}
    s=\frac{g^4\mu_B}{F_1}=\frac{g^4\mu_B}{\frac{2k_BT_A^2}{15cT_0}}=\frac{T_0}{\left(248 \frac{Oe}{K}\right) T_A^2}
    \label{SE1}
\end{equation}
\begin{equation}
    \mu_s^2=20 \frac{T_0}{T_A} C_{4/3}\left(\frac{T_C}{T_0}\right)^{4/3} \approx 20\frac{T_C^{4/3}}{T_AT_0^{1/3}},
    \label{SE2}
\end{equation}
when using $g=2$, $C_{4/3}\approx 1$, $c=1/2$ (as done in Ref. [\citenum{Takahashi}]), $\mu_B=9.27\times 10^{-21}\ \text{emu}$ and $k_B=1.38\times 10^{-16}\ \text{erg}/\text{K}$. $T_0$ is a spectral parameter associated with the width of the frequency dependence of the generalized susceptibility at the zone boundary, and $T_A$ is associated to the width of its $\mathbf{q}$ dependence around $\mathbf{q}=0$. The system of equations given by (\ref{SE1}) and (\ref{SE2}) can be solved for $T_0$ and $T_A$, yielding
\begin{equation}
    T_0=\left(\frac{20}{\mu_s^2}\right)^{6/5}\left(248\frac{\text{Oe}}{\text{K}}\ s\right)^{3/5} T_C^{8/5},
\end{equation}
\begin{equation}
    T_A=\left(\frac{20}{\mu_s^2}\right)^{3/5}\left(248\frac{\text{Oe}}{\text{K}}\ s\right)^{-1/5}T_C^{4/5}.
\end{equation}

Taking as an example the case of $x=0.98$ shown in Fig. \ref{fig:Arrot}, for the red curve corresponding to 2 K, the slope is $s=3.86(1)\times10^{-9}\ \mu_B^3/\text{Oe}$ and the intercept is $\mu_s^2=0.914(4)\times10^{-3}\ \mu_B^2$, leading to $T_0=800(100)\ \text{K}$.

In order to calculate $\mu_s$ and $T_0$ for $x=0.96$ shown in Fig. \ref{fig:arrot_PR598}, a linear fit was performed for the range of fields above the metamagnetic transition for $x=0.96$ at $T=2\ \text{K}$ (dashed line). This was done in order to be consistent with the range of $H/M$ in which the fits were done for the Deguchi-Takahashi analysis on the other samples, including those with AFM ground states. 

As mentioned in the Discussion, $\tilde{T}_0$ was calculated according to
\begin{equation}
    \tilde{T}_0=\left(\frac{20T_{mag}}{\mu_s^2}\right)^{6/5}\left(248\frac{\text{Oe}}{\text{K}}\ s\right)^{3/5},
\end{equation}
in order to obtain the values included in Table \ref{tab:tilde}, which were used for the plot in the inset of Fig. \ref{fig:Takahashi_FM}.

\begin{table}[htbp]
  \centering
  \setlength{\tabcolsep}{1.5pt}
  \renewcommand{\arraystretch}{1.75}
    \begin{tabular}{|c|c c c|c c c|}
    \hline
    $x$     &    $T_{mag}$   &  $\tilde{T}_0$  & $T_{mag}/\tilde{T}_0$   & $\mu_{eff}$  & $\mu_{s}$ & $\mu_{eff}/\mu_{s}$\\
         &    (K)   &  (K)  &    & $(\frac{\mu_B}{TM})$  & $(\frac{\mu_B}{TM})$ & \\
    \hline
    0.65     &   8(1)  & 690(20)  &  0.012(2)  &  0.897  &  0.0357(4)  & 25.3(4)\\
    0.74     &    16(1)   &  540(20)  & 0.030(3)  & 1.053  & 0.074(1) & 14.3(3)\\
    0.82     &   22(1)   &  640(20)  & 0.034(3)  & 1.177  & 0.089(1) & 13.3(3)\\
    0.85     &    25(1)   &  620(50) &  0.040(5)
    & 1.21  & 0.101(3) & 12.1(6)\\
    0.87     &    30(1)   &  400(20) &  0.075(6)  & 1.31  & 0.148(3) & 9.0(3)\\
    0.88     &    36(1)   &  490(20)  & 0.073(4)  & 1.151  & 0.150(1) & 7.8(2)\\
    0.90     &    36(1) &  480(20) & 0.075(5) & 1.22  & 0.150(4) & 8.2(3)\\
    0.91     &    35(3)   &  630(10) &  0.056(3)  & 1.208  & 0.144(1) & 8.5(1)\\
    0.94     &    27(4)   &  1050(40) &  0.026(2)  & 1.23  & 0.092(1) & 13.0(3)\\
    0.95     &    $-$   &  $-$ & $-$   & 1.322  & 0.071(1) & 18.9(5)\\
    0.96     &    6.4(6)   &  350(10) &  0.019(4)  & 1.36  & 0.0501(6) & 27.4(7)\\
    0.98     &    6.6(4)   &  800(100) &  0.008(3)  & 1.37  & 0.030(3) & 46(5)\\
    0.98     &    4.3(7)   &  1330(80) &  0.0032(9)  & 1.54  & 0.0145(3) & 107(3)\\

          \hline
    \end{tabular}%
    \caption{\footnotesize Parameters used to build the modified Takahashi-Deguchi plot on the inset of Fig. \ref{fig:Takahashi_FM}.}
  \label{tab:tilde}%
\end{table}%

\label{sec:Appendix}

\bibliographystyle{apsrev4-1}
\bibliography{CollapsedTet}

\end{document}